\documentclass[%
 reprint,
superscriptaddress,
nobibnotes,
 amsmath,amssymb,
 aps,
prb,
]{revtex4-2}

\usepackage[colorlinks=true,linkcolor=blue,urlcolor=blue,citecolor=blue]{hyperref}
\usepackage{graphicx}% Include figure files
\usepackage{dcolumn}% Align table columns on decimal point
\usepackage{bm}% bold math
\usepackage{multirow}
\usepackage{float}
\usepackage{setspace}
\usepackage{gensymb}

\begin{document}

\title{Quantum spin Hall effect protected by spin U(1) quasisymmetry}

% Force line breaks with \\

\author{Lu Liu}
\thanks{Lu Liu and Yuntian Liu contributed equally to this work.}
\affiliation{Department of Physics and Shenzhen Institute for Quantum Science and Engineering (SIQSE), Southern University of Science and Technology, Shenzhen 518055, China}
 \affiliation{Laboratory for Computational Physical Sciences (MOE),
 State Key Laboratory of Surface Physics, and Department of Physics,
  Fudan University, Shanghai 200433, China}

\author{Yuntian Liu}
\thanks{Lu Liu and Yuntian Liu contributed equally to this work.}
\affiliation{Department of Physics and Shenzhen Institute for Quantum Science and Engineering (SIQSE), Southern University of Science and Technology, Shenzhen 518055, China}

\author{Jiayu Li}
\affiliation{Department of Physics and Shenzhen Institute for Quantum Science and Engineering (SIQSE), Southern University of Science and Technology, Shenzhen 518055, China}

\author{Hua Wu}
\affiliation{Laboratory for Computational Physical Sciences (MOE),
 State Key Laboratory of Surface Physics, and Department of Physics,
 Fudan University, Shanghai 200433, China}
 \affiliation{Shanghai Qi Zhi Institute, Shanghai 200232, China}
\affiliation{Collaborative Innovation Center of Advanced Microstructures,
 Nanjing 210093, China}

\author{Qihang Liu}
\email{Corresponding author. liuqh@sustech.edu.cn}
\affiliation{Department of Physics and Shenzhen Institute for Quantum Science and Engineering (SIQSE), Southern University of Science and Technology, Shenzhen 518055, China}
\affiliation{Guangdong Provincial Key Laboratory of Computational Science and Material Design, Southern University of Science and Technology, Shenzhen 518055, China}

\date{\today}

\begin{abstract}
Quantum spin Hall (QSH) insulators, with their unique helical edge states where counterpropagating edge channels possess opposite spins, have attracted broad interest across various fields. While the exact quantization of spin Hall conductance (SHC) is elusive in realistic materials due to intrinsic spin mixing effects, the near-quantization, as a compromised definition of the QSH effect, cannot be captured by rigorous topological invariants. In this Letter, we present a universal symmetry indicator for diagnosing the QSH effect in realistic materials, termed spin U(1) quasisymmetry. Such a symmetry eliminates the first-order spin-mixing perturbation and thus protects the near-quantization of SHC, applicable to time-reversal-preserved cases with either $Z_{2}=1$ or $Z_{2}=0$, as well as time-reversal-broken scenarios. We propose that spin U(1) quasisymmetry is hidden in the subspace spanned by the doublets with unquenched orbital momentum and emerges when SOC is present, which can be realized in 19 crystallographic point groups. Our theory is applied to identify previously overlooked QSH phases such as time-reversal-preserved even spin Chern phase and time-reversal-broken phase, as exemplified by twisted bilayer transition metal dichalcogenides, monolayer RuBr$_{3}$ and monolayer FeSe. Our work provides a comprehensive symmetry-based framework for understanding the QSH effect and significantly expands the material pool for the screening of exemplary material candidates.
\end{abstract}

\maketitle

$Introduction.\quad$Two-dimensional (2D) quantum spin Hall (QSH) insulators manifest a plateau of spin Hall conductance (SHC) with a nearly quantized value inside the energy gap of the bulk \cite{QSH_gra,QSH_SCZ}. They have triggered the recent prosperity of topological phases of matter and topological electronics owing to their promising potential dissipationless spin transports and broad impact across diverse fields \cite{ZSC_Science2006,HgTe_exp1,TI_RMP2010,TI_RMP2011,InAs_exp,SOT,SC_1,transistor,thermo,nonlinear}. Conventional QSH insulators are diagnosed by a nontrivial $Z_{2}$ index as the symmetry indicator, protected by time-reversal symmetry (TRS) \cite{Z2_2005}, and are expected to exhibit quantized SHC. However, it has come to light recently that certain $Z_{2}=1$ QSH insulators manifest significant deviations from the expected quantized value $ne/2\pi$ with $n$ being an integer and $e$ denoting the electron charge \cite{WTe2_NP2,WTe2_science,WTe2_prx,SHC_prb2019}, with the underlying reasons being discussed through both intrinsic \cite{WTe2_prx,SHC_prb2019,WTe2_canted,WTe2_nl} and extrinsic \cite{ex1,ex2,ex3,ex4} mechanisms. More intriguingly, experiments have observed the near-double-quantized conductance in twisted bilayer WSe$_{2}$ and  MoTe$_{2}$ within the $Z_{2}=0$ regime \cite{twist_arxiv1,twist_arxiv2}. In addition, recent observations also reveal the helical edge state in antiferromagnetic FeSe monolayer \cite{FeSe_NM,FeSe_nl}. These findings indicate that the QSH effects can persist even in the absence of a nontrivial $Z_{2}$.

Generically, quantized SHC, a hallmark of the QSH phase, is calculated by $\sigma_{xy}^{S}=C_{S}(\frac{e}{2\pi})$ \cite{Haldane2006}, where the spin Chern number $C_{S}$, defined by the difference between the Chern numbers of spin-up and spin-down channels, serves as a topological invariant \cite{Haldane2006,SCN_robust,Sheng_PRL2011}. However, it is notable that the ``spin'' in $C_{S}$ is uniquely defined for each wavevector $\bm{k}$ in the Brillouin zone if spin-mixing interactions exist. When considering the collective behavior of all the $\bm{k}$ points, e.g., SHC, the real spin is no longer a good quantum number \cite{Z2_2005,Haldane2005,Haldane2006,SCN_robust,Sheng_PRL2011}. Therefore, from an intrinsic perspective, SHC becomes exactly quantized only when the real-spin-component $S_{z}$ is preserved, corresponding to a U(1) spin rotation symmetry  \cite{Z2_2005,Haldane2005,Haldane2006,wenxg}. This leads to a crucial question that cannot be addressed by the topological invariants $Z_{2}$ and $C_{S}$: how can the near-quantized SHC plateau emerging from the bulk gap of a 2D material, which can be a compromised definition of QSH effect, be protected and understood by symmetry? Several studies have sought to unravel the factors influencing QSH effect. For instance, Rashba SOC has been demonstrated to be destructive to QSH effect, arising from the breakdown of in-plane mirror symmetry \cite{Z2_2005,QSH_gra,Haldane2005}. Moreover, high in-plane symmetry has been suggested to enhance the potential for near-quantization feature via comparisons among QSH insulators with hexagonal, square, and rectangular lattices \cite{SHC_prb2019}. Furthermore, the dependence of SHC values on the spin axis has also been investigated \cite{WTe2_prx,WTe2_canted,WTe2_nl}. Yet, an effective symmetry indicator for the QSH effects in realistic materials remains elusive.

In this Letter, we show that besides a nontrivial $C_{S}$, there is a previous overlooked symmetry indicator for the diagnosis of QSH effect, namely spin U(1) quasisymmetry. The significance of such an approximate symmetry is to eliminate the first-order spin-mixing interactions, thereby protecting the near-quantization of SHC. We propose that the spin U(1) quasisymmetry is inherently hidden in the subspace formed by the states possessing unquenched orbital angular momentum $l_{z}$, and emerges when SOC is present. Such states are characterized by 1D complex irreducible representations (irreps) and 2D irreps supported in 19 crystallographic point groups. Interestingly, we identify an even spin Chern (ESC) phase with a trivial $Z_{2}$ index but a nontrivial $C_{S}$ and spin U(1) quasisymmetry as an ideal platform to realize high near-quantized SHC. By symmetry analysis and first-principles calculations, we provide design principles for such ESC insulators, exemplified by twisted bilayer transition metal dichalcogenides (TMDs) and RuBr$_{3}$ monolayer. Furthermore, we also apply our theory to antiferromagnetic FeSe monolayer, demonstrating that near-quantized SHC could appear even in TRS-broken systems with spin U(1) quasisymmetry.

$Spin$ U(1) $quasisymmetry.\quad$ Quasisymmetry was originally introduced to account for the near band-degeneracies and the resultant large berry curvature in CoSi \cite{quasi_np,quasi_prb}. Later, a generic theory on quasisymmetry was developed, expanding the conventional group theory to address questions regarding large-or-small splittings \cite{quasi_JY}. Specifically, quasisymmetry describes the hidden symmetry within a specific degenerate eigensubspace under the unperturbed Hamiltonian $H_{0}$, thereby forming an enlarged group beyond the symmetry group of $H_{0}$ (denoted as $\mathcal{G}_{H_{0}}$) \cite{quasi_JY}. Importantly, the presence of quasisymmetry constrains the symmetry-lowering interactions $H^{\prime}$ to operate only as a second-order effect. Considering the subspace formed by the spin eigenstates, $\left | \uparrow \right \rangle$ and $\left | \downarrow \right \rangle$, it is invariant under spin U(1) symmetry operator $e^{i\theta \sigma_{z}}$. When adding a spin-mixing perturbation $H^{\prime}$, the term $\left\langle \uparrow \right | H^{\prime} \left |\downarrow \right\rangle$, characterizing the first-order perturbation breaking the spin U(1) symmetry, is allowed to be non-zero if it remains invariant under all operations in $\mathcal{G}_{H_{0}}$. However, the inherent spin U(1) symmetry within the spin eigensubspace gives that 
\begin{equation}
\label{equation1}
\left\langle \uparrow \right | H^{\prime} \left |\downarrow \right\rangle \overset{e^{i\theta \sigma_{z}}} {\longrightarrow}
\left\langle \uparrow \right | e^{i\theta \sigma_{z}} H^{\prime} (e^{i\theta \sigma_{z}})^{-1}\left |\downarrow \right\rangle =e^{i2\theta}\left\langle \uparrow \right | H^{\prime}\left |\downarrow \right\rangle.
\end{equation}
The presence of the phase factor $e^{i2\theta}$ enforces 
\begin{equation}
\label{equation2}
\left\langle \uparrow \right | H^{\prime} \left |\downarrow \right\rangle=0, 
\end{equation}
suggesting that the spin-mixing perturbation $H^{\prime}$ acts as at least a second order. Therefore, while excluded from $\mathcal{G}_{H_{0}}$, the spin U(1) symmetry exists as a quasisymmetry within the subspace spanned by $\{\left | \uparrow \right \rangle$, $\left | \downarrow \right \rangle\}$, and significantly, it eliminates the first-order spin-mixing perturbation, which is a dominant detriment to the quantization in QSH phase. 

\begin{figure}[t]
  \centering
\includegraphics[width=8.5cm]{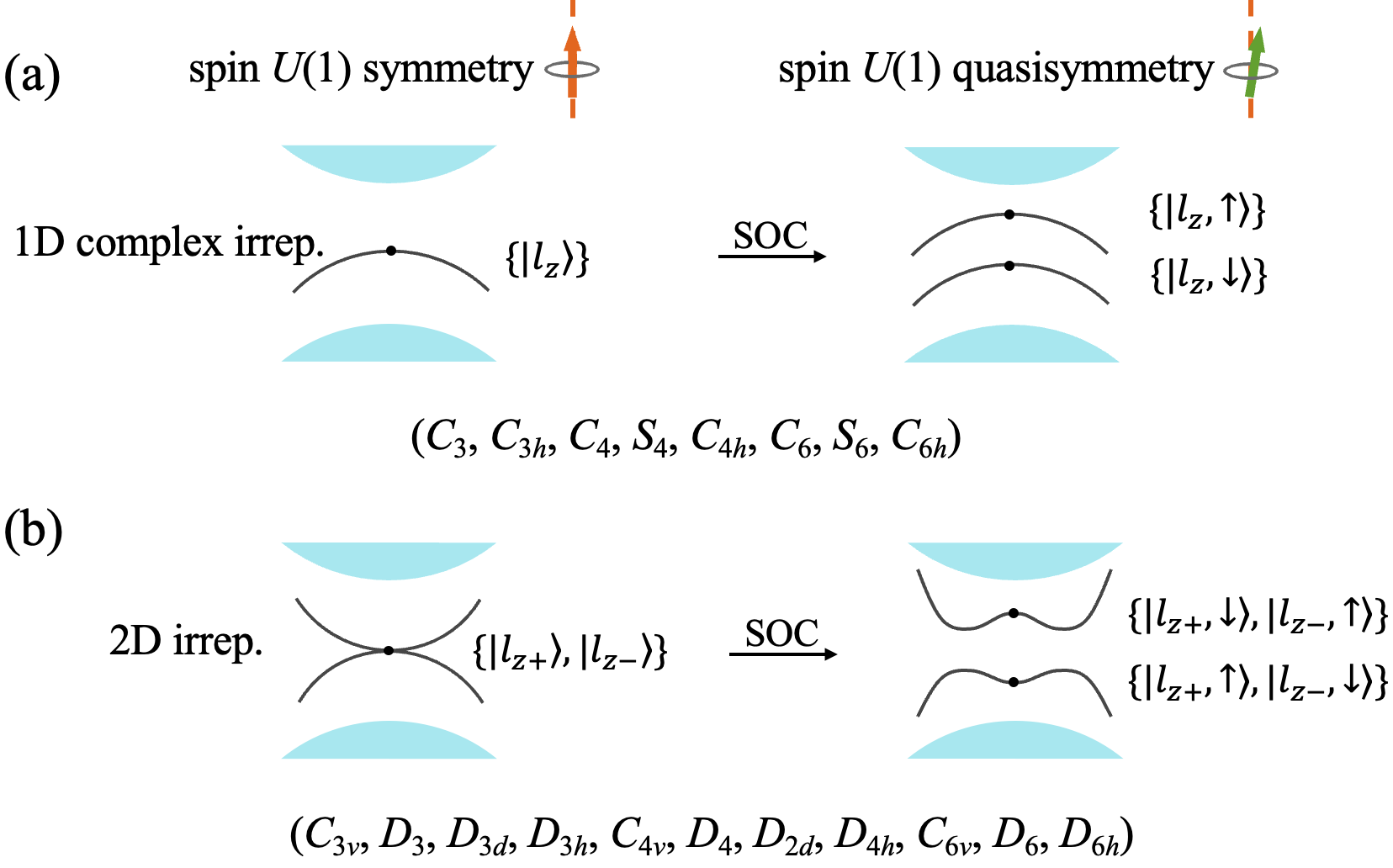}
  \caption{Schematic of the spin U(1) quasisymmetry and its design principles: Spin U(1) quasisymmetry is hidden in the subspace formed by the states with nonzero $l_{z}$, which is allowed by (a) the spin doublet characterized by 1D complex irreps and (b) the orbital doublet characterized by 2D irreps, and emerges when SOC is present.}
  \label{fig:1}
\end{figure}

An important subsequent step is to identify the conditions that cause spin U(1) quasisymmetry in real materials. The key point is to ensure that within an eigensubspace that is invariant under $e^{i\theta \sigma_{z}}$, the spin-mixing term merely acts as a perturbation relative to the spin-preserving term. We propose that such conditions are satisfied by utilizing the states with unquenched orbital angular momentum $l_{z}$. The simplest case is a spin-degenerate band in nonmagnetic materials. As shown in Fig. \ref{fig:1}(a), in the spin doublet $\{\left| l_{z},\uparrow \right \rangle, \left| l_{z},\downarrow \right \rangle \}$, the spin-preserving SOC matrix elements $\left\langle l_{z}, \uparrow \right | \sigma_{z} L_{z} \left | l_{z},\uparrow \right\rangle =l_{z}$ and $\left\langle l_{z}, \downarrow \right | \sigma_{z} L_{z} \left | l_{z},\downarrow \right\rangle =-l_{z}$ are non-vanishing. In contrast, the spin-mixing SOC elements $\left\langle l_{z}, \uparrow \right | \sigma_{x} L_{x} + \sigma_{y} L_{y} \left | l_{z},\downarrow \right\rangle$ are zero unless involving remote bands to invoke a second-order effect. In other words, when involving the bands beyond the subspace of $\{\left| l_{z},\uparrow \right \rangle, \left| l_{z},\downarrow \right \rangle \}$, the spin-mixing interaction $H^{\prime}$ can be introduced but emerge as a perturbation compared to the spin-preserving interaction $H_{0}$, see more details in the Sec. S1 of the Supplemental Material S1 \cite{[{See Supplemental Materials for notes regarding second-order spin-mixing perturbation, continuum model Hamiltonian, DFT calculations, and effective Hamiltonian of the magnetic point group $6^{\prime}mm^{\prime}$}]SM} (see also references \cite{SI1,SI3,SI4,SI5,SI6,SI8,SI9,SI10,SI11,SI13,SI14,SI15,SI16,SI21,SI22,SI23,SI24,SI25,SI26,SI28} therein). Upon projecting the Hamiltonian onto the subspace of $\{\left| l_{z},\uparrow \right \rangle, \left| l_{z},\downarrow \right \rangle \}$, the unperturbed Hamiltonian $H_{0}$ and the perturbed Hamiltonian $H^{\prime}$ around $\bm{q}$ can be generically written as $H_{0}\sim \sigma_{z} [ 1 + h_{0}(\bm{k})]$ and $H^{\prime}\sim \sigma_{x,y} h^{\prime}(\bm{k})$, respectively, where $h_{0}(\bm{k})$ and $h^{\prime}(\bm{k})$ are exactly zero at $\bm{q}$. Indeed, under the action of spin U(1) symmetry operator $e^{i\theta \sigma_{z}}$, $H_{0}$ remains invariant, whereas $H^{\prime}$ does not. Figure \ref{fig:1}(a) represents the bands with unquenched $l_{z}$ at non time-reversal-invariant momenta, such as the top valance bands near the $K$ valley in $H$-TMDs \cite{valley_yao}.

The spin-preserving SOC also dominates when the subspace is spanned by an orbital doublet, as depicted in Fig. \ref{fig:1}(b). Therefore, with SOC the subspace is formed by $\{\left | \uparrow \right \rangle$, $\left | \downarrow \right \rangle\}  \otimes \{\left | l_{z+} \right \rangle$, $\left | l_{z-} \right \rangle\} $, where $l_{z\pm}$ denotes opposite orbital angular momentum. In this case, the unperturbed Hamiltonian $H_{0}$ and the perturbed Hamiltonian $H^{\prime}$ take the form of $H_{0}\sim \sigma_{z}\otimes [\tau_{z} + h_{0}(\bm{k})\tau_{x,y,z}]$ and $H^{\prime}\sim \sigma_{x,y}\otimes h^{\prime}(\bm{k})\tau_{x,y,z}$, respectively, where $\sigma$/$\tau$ denotes Pauli matrices for spin/orbital degrees of freedom. Therefore, spin U(1) symmetry emerges as a quasisymmetry of the subspace spanned by the states with nonzero $l_{z}$. When the corresponding energy bands manifest nontrivial $C_{S}$, their inherent spin U(1) quasisymmetry protects the near-quantized SHC within the bulk insulating gap by eliminating the detrimental first-order spin-mixing perturbation. Note that in contrast to the orbital quasisymmetry used to protect small band gaps \cite{quasi_np,quasi_prb,quasi_JY,aozhang}, the spin U(1) quasisymmetry is instead related to large band gaps opened by SOC, which, as a bonus, help against thermal fluctuations in QSH insulators.

Regarding the symmetry requirement, the spin doublet carrying nonzero $l_{z}$ in Fig. \ref{fig:1}(a) is characterized by 1D complex irreps. This condition is satisfied by eight crystallographic point groups: $(C_{3}, C_{3h}, C_{4}, S_{4}, C_{4h}, C_{6}, S_{6}, C_{6h})$. The orbital doublet shown in Fig. \ref{fig:1}(b) is supported by 11 point groups $(C_{3v}, D_{3}, D_{3d}, D_{3h}, C_{4v}, D_{4}, D_{2d}, D_{4h}, C_{6v},$ $D_{6}, D_{6h})$ with 2D irreps and the abovementioned eight groups with conjugate 1D complex irreps in TRS-preserved systems. Consequently, there are in total 19 crystallographic point groups supporting the states with nonzero $l_{z}$ and thus spin U(1) quasisymmetry. Note that a similar approach can be directly extended to magnetic materials, which will be discussed later.

It is worth noting that our discussions about spin U(1) quasisymmetry is not confined to $Z_{2}=1$ or $Z_{2}=0$ cases. In fact, the spin-mixing terms induced by symmetry breaking could also weaken the SHC in the $Z_{2}=1$ conventional QSH insulators, as demonstrated in previous studies \cite{Z2_2005,QSH_gra,Haldane2005,SHC_prb2019}. Therefore, for these $Z_{2}=1$ systems, spin U(1) quasisymmetry which eliminates the first-order spin-mixing perturbation still plays a crucial role for protecting the near-quantization. In the following section, we present two types of $Z_{2}=0$ spin Chern insulators with spin U(1) quasisymmetry and show that such a topological phase manifests a high and near-quantized SHC plateau, rendering an ideal platform for observing QSH effects.

$Twisted$ $bilayer$ $TMDs$: $Spin$ $doublet.\quad$ We next provide a realistic example of ESC insulators with spin U(1) quasisymmetry promoted by the spin doublet [Fig. \ref{fig:1}(a)]: twisted bilayer TMDs, in which the moiré valence bands are formed from the states at the $\pm K$ valleys of two $H$-TMD monolayers. Although individual $H$-TMD monolayers do not exhibit QSH effect \cite{OHE}, moiré bands in twisted bilayers have been theoretically predicted and experimentally confirmed to manifest a nontrivial ESC phase \cite{twist_MCD,twist_fu, twist_arxiv1,twist_arxiv2}. Here we emphasize that while the nontrivial $C_{S}$ is caused by the moiré structure, the observed QSH effects also benefit from the spin U(1) quasisymmetry hidden in the electronic structure of each $H$-TMD monolayer. For $H$-TMD monolayers, the topmost valence states at the $+K$-valley with the little point group $C_{3h}$ consisting of spin doublet $\{\left| l_{z}=+2,\uparrow \right \rangle, \left| l_{z}=+2,\downarrow \right \rangle \}$ \cite{valley_yao} when SOC is absent, in excellent agreement with the symmetry requirement of Fig. \ref{fig:1}(a). To see the role of spin U(1) quasisymmetry in twisted bilayer TMDs, we adopt the continuum model Hamiltonian at the $+K$ valley and consider both spin-preserving and spin-mixing effects by including the $\left| l_{z}=+2,\uparrow \right \rangle$ and $\left| l_{ z}=+2,\downarrow \right \rangle$ states \cite{SM}. When SOC emerges, the spin degeneracy at $+K$ is lifted, resulting in a strong spin-preserving SOC gap ($\sim$ 200 meV \cite{twist_MCD}) between the $\left| l_{z}=+2,\uparrow \right \rangle$ band  and the $\left| l_{z}=+2,\downarrow \right \rangle$ band.

\begin{figure}[t]
  \centering
\includegraphics[width=8.5cm]{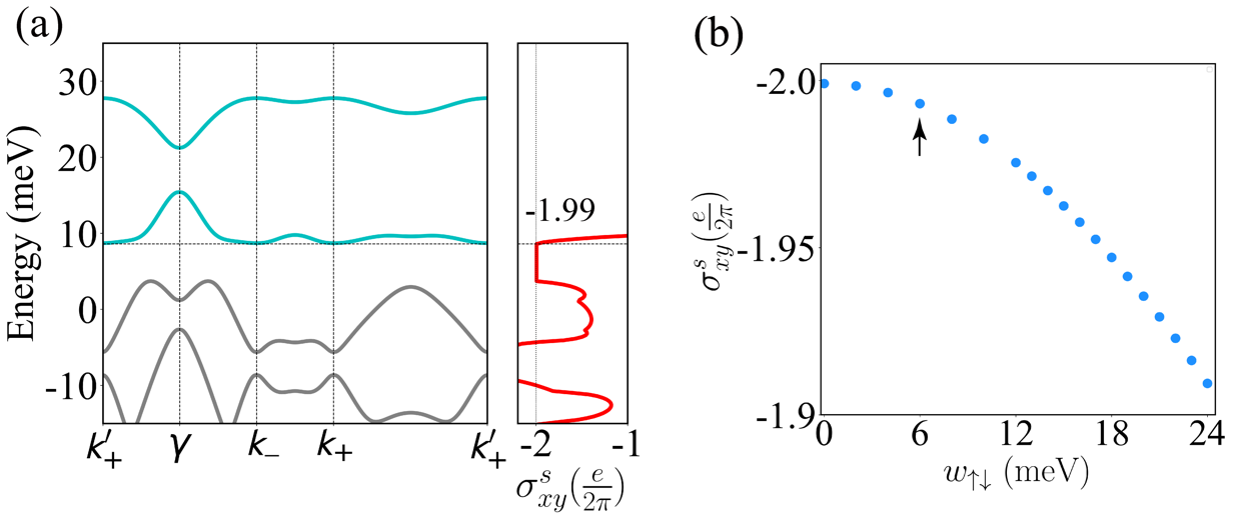}
  \caption{(a) Moiré band structure at a twist angle of $2.5\degree$. The top two bands are indicated in green curves. The energy dependence of SHC is plotted in the right panel. (b) SHC values as a function of the spin-mixing strength $w_{\uparrow\downarrow}$. The arrow marks the $w_{\uparrow\downarrow}$ value of 6 meV used in (a).}
  \label{fig:2}
\end{figure}

We calculate the moiré band structure at a twist angle of $2.5\degree$. As shown in Fig. \ref{fig:2}(a), the top two bands are well separated from the third band. By integrating the Berry curvature within the first moiré Brillouin zone, we find that the first two moiré bands at the $+K$ valley possess a nontrivial Chern number of $-2$. Due to spin-valley locking in TMDs, the $-K$ valley must carry a Chern number of $2$. Therefore, the twisted bilayer is in an ESC phase with $|C_{S}|=2$. Moreover, our calculations on SHC reveal a plateau within the energy gap between the second and third bands corresponding to the moiré hole filling factor $\nu=4$, as shown in the right panel of Fig. \ref{fig:2}(a). Notably, the SHC value of $-$1.99 closely approaches the quantized value of $-2$. This is attributed to the essential role of spin U(1) quasisymmetry eliminating the first-order perturbation of spin-mixing interaction, i.e., Eq. (\ref{equation2}). Furthermore, we show that due to that spin-mixing effects act at least a second-order perturbation, even increasing the spin-mixing strength to larger than 20 meV, the SHC value remains high and nearly quantized to $-2$, as shown in Fig. \ref{fig:2}(b). Therefore, we conclude that $|C_{S}|=2$ and spin U(1) quasisymmetry, which originates from the moiré band renormalization and the spin doublets with unquenched $l_{z}$ for a monolayer, respectively, are both indispensable factors to ensure the near-quantized SHC successfully observed in twisted bilayer TMDs \cite{twist_arxiv1,twist_arxiv2}. Note that spin U(1) quasisymmetry remains intact under an out-of-plane magnetic field, as it merely introduces a spin-preserving term $B_{z}\sigma_{z}$ into $H_{0}$. However, an in-plane magnetic field with the form of $B_{x, y}\sigma_{x, y}$ disrupts the doubly degenerate subspace $\{\left| l_{z}=+2,\uparrow \right \rangle, \left| l_{z}=+2,\downarrow \right \rangle \}$ and thus breaks the spin U(1) quasisymmetry. This aligns with the experimental findings where the conductance remains stable under out-of-plane magnetic fields but decreases under in-plane fields \cite{twist_arxiv1,twist_arxiv2}.

\begin{figure}[t]
\includegraphics[width=8.5cm]{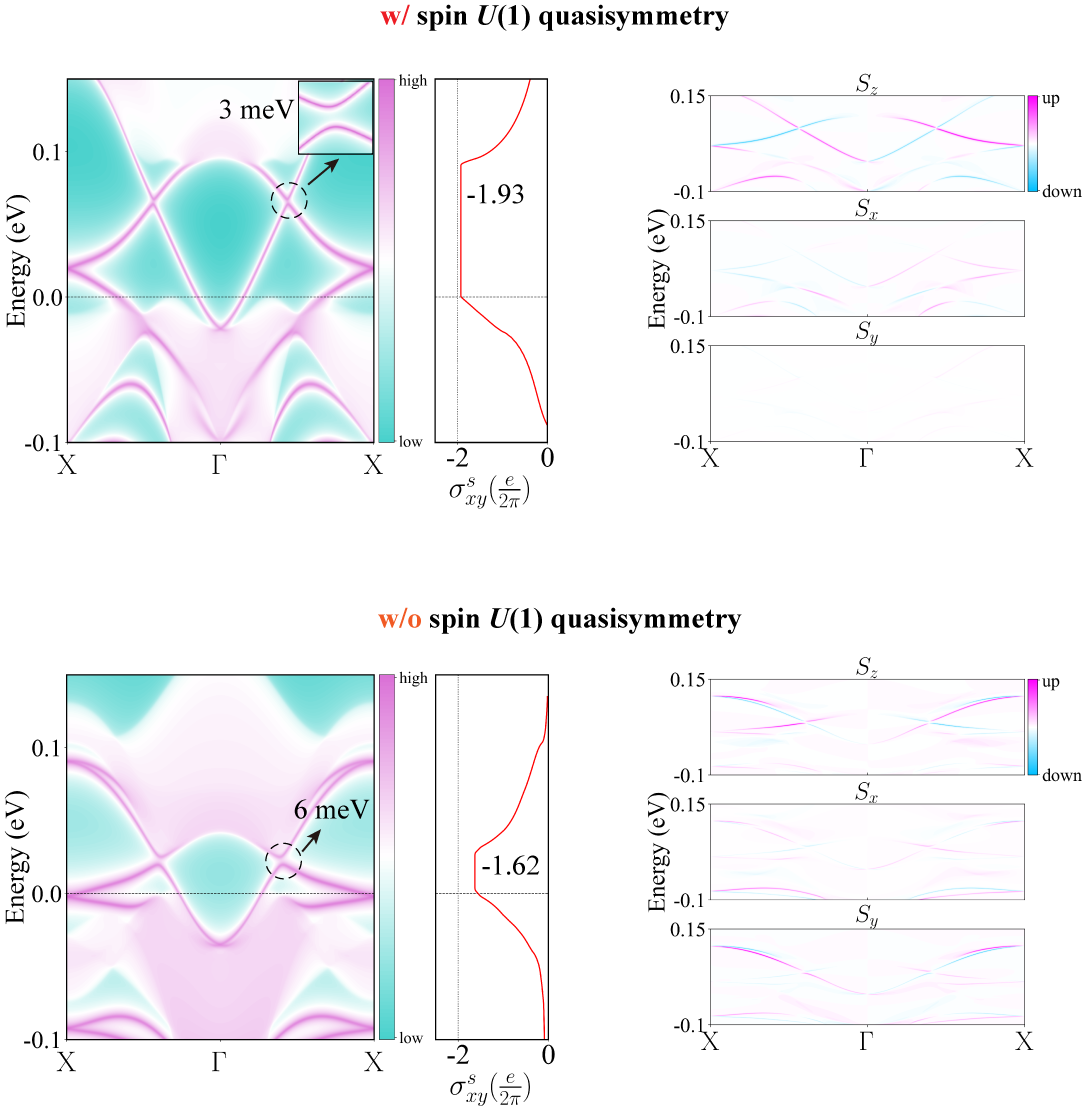}
  \caption{(a, c) The edge states and SHC, and (b, d) spin components of the edge states in RuBr$_{3}$ monolayer. The bands with sharper purple are more localized on the boundary. The Fermi level is indicated as the dashed lines. The undistorted structure is adopted in (a-b), in which spin U(1) quasisymmetry is present in the low-energy states. In contrast, the crystal structure in (c-d) undergoes a deformation to a low-symmetry phase under a uniaxial tensile strain, resulting in the absence of spin U(1) quasisymmetry.}
  \label{fig:3}
\end{figure}

RuBr$_{3}$ $monolayer$: $Orbital$ $doublet.\quad$Next, we introduce another example of ESC insulators with spin U(1) quasisymmetry supported by orbital doublet: the RuBr$_{3}$ monolayer. The monolayer has a space group $P\bar{3}1m$, with the little point group $D_{3d}$ at the $\Gamma$ point, as shown in Fig. \ref{fig:1}(b). The orbital doublet $\{\left| e^{\prime}_{+} \right \rangle, \left| e^{\prime}_{-} \right \rangle \}$ governs the low-energy physics of RuBr$_{3}$ \cite{lliu}. This degeneracy is lifted by introducing SOC effects, and a bulk gap of about 100 meV is opened (see Figs. S3 and S4 \cite{SM}). Notably, the onsite first-order spin-preserving SOC contributes a significant gap of larger than 300 meV at the $\Gamma$ point. To investigate the topological nature of the RuBr$_{3}$ monolayer, we first calculate the $Z_{2}$ index by computing the parity eigenvalues of valence bands at the time-reversal-invariant momenta \cite{FuKane_Z2}. The same parity at $\Gamma$ and M results in $Z_{2}=0$, suggesting that RuBr$_{3}$ is a $Z_{2}$ trivial insulator. Interestingly, despite RuBr$_{3}$ belongs to the $Z_{2}=0$ phase, its nontrivial topological features are evident in the edge spectrum and SHC. As shown in Figs. \ref{fig:3}(a) and (b), RuBr$_{3}$ exhibits two pairs of helical edge states and two Dirac-like edge crossings with tiny gaps of 3 meV. Furthermore, the SHC plateau of RuBr$_{3}$ persists within the bulk gap, and its value of 1.93 closely approximates the quantized value of 2. Through adiabatically restoring the spin $U(1)$ symmetry \cite{SCN_robust,Sheng_PRL2011,SCN_exp} and calculating the $C_{S}$, we confirm that the RuBr$_{3}$ monolayer is in an ESC phase characterized by $C_{S}=-2$.

Here we construct an effective model Hamiltonian around the $\Gamma$ point to capture the topological nature and the quasisymmetry of RuBr$_{3}$. The generators of the little point group $D_{3d}$ at the $\Gamma$ point are threefold rotation symmetry $C_{3z}$ along the $z$ axis, twofold rotation symmetry $C_{2y}$ along the $y$ axis, and space inversion symmetry $I$. In the basis of $\{|e^{\prime}_{+},\uparrow\rangle, |e^{\prime}_{-},\uparrow\rangle, |e^{\prime}_{+},\downarrow\rangle, |e^{\prime}_{-},\downarrow\rangle\}$, the representation matrices of the symmetry operations are given by $C_{3z}=e^{i\frac{\pi}{3}\sigma_{z}}\otimes e^{-i\frac{2\pi}{3}\tau_{z}}$, $C_{2y}=e^{i\frac{\pi}{2}\sigma_{y}}\otimes\tau_{x}$, $I=\mathbb{I}_{2 \times 2}\otimes -\mathbb{I}_{2 \times 2}$, and TRS $T=\mathcal{K}\cdot i \sigma_{y}  \otimes \tau_{x}$, where $\mathcal{K}$ is the complex conjugation operator and $\mathbb{I}_{2 \times 2}$ is a $2 \times 2$ identity matrix. By imposing those symmetries, we derive the generic form of the effective Hamiltonian $H(\bm{k})$ as follows:
\begin{equation}
\begin{aligned}
\label{equation3}
H(\bm{k})=&H_{0}(\bm{k})+ H^{\prime}(\bm{k}) \\
H_{0}(\bm{k})=&\epsilon_{0}(\bm{k}) \mathbb{I}_{4 \times 4} + C[(k_{x}^{2}-k_{y}^{2})\sigma_{0}\otimes\tau_{x}+2k_{x}k_{y}\sigma_{0}\otimes\tau_{y}]\\
& + D (k_{x}^{2}+k_{y}^{2})\sigma_{z}\otimes\tau_{z}+E \sigma_{z}\otimes\tau_{z}\\
H^{\prime}(\bm{k})=&F[(k_{x}^{2}-k_{y}^{2})\sigma_{x}\otimes\tau_{z}+2k_{x}k_{y}\sigma_{y}\otimes\tau_{z}]\\
\end{aligned}
\end{equation}
where $\epsilon_{0}(\bm{k})=A-B(k_{x}^{2}+k_{y}^{2})$. For a negative $E$, this model describes the system in a nontrivial ESC phase with $C_{S}=-2$. Notably, within the subspace formed by orbital doublet and electron spin, spin-mixing interaction is merely a perturbation compared to the dominant spin-preserving SOC effect. With respecting the time-reversal and space inversion symmetries in the bulk phase, the spin-mixing $H^{\prime}$ in Eq. (\ref{equation3}) does not lift the spin degeneracy but breaks the spin U(1) symmetry. However, the spin U(1) quasisymmetry embedded in the eigenspace of the unperturbed Hamiltonian $H_{0}$ in Eq. (\ref{equation3}) eliminates the first-order perturbation effect of $H^{\prime}$ [Eq. (\ref{equation2})], thereby protecting the QSH effect. This is verified by our density functional theory (DFT) calculated results \cite{SM}: (i) a relatively small edge gap of 3 meV induced by second-order spin-mixing perturbation; (ii) the predominance of the spin-preserved $S_{z}$ component in the edge states; (iii) a persistent SHC plateau within the bulk gap, with a value of $-$1.93, closely approximating the quantized value of $-$2, as shown in Figs. \ref{fig:3}(a) and (b).

To further validate our theory, we apply a uniaxial tensile strain to break the spin $U(1)$ quasisymmetry in RuBr$_{3}$. Specifically, the strain reduces the little group of the $\Gamma$ point from $D_{3d}$ to $C_{i}$, eliminating the orbital doublet and the spin $U(1)$ quasisymmetry. As illustrated by the comparisons between Figs. \ref{fig:3}(a)-\ref{fig:3}(d), the breakdown of spin $U(1)$ quasisymmetry leads to a considerable decrease in SHC values from $-$1.93 to $-$1.62, accompanied by an increase in the edge gap and enhanced spin-mixing $S_{x,y}$ components in the edge states. Consequently, the essential role of spin $U(1)$ quasisymmetry in protecting the near-quantized SHC is verified. In addition to those characteristics in edge spectra, complementary evidences concerning the $k$-resolved spin Berry curvature and spin gap of all occupied bulk bands are provided in Fig. S6 \cite{SM}. Specifically, with the top valence bands predominantly contributing the spin Berry curvature, which is concentrated around the $\Gamma$ point, the presence of spin U(1) quasisymmetry results in larger spin gaps at the $\Gamma$ point and throughout the Brillouin zone compared to the scenario without spin U(1) quasisymmetry, signifying the weakened spin-mixing interaction and its suppressed disruption to SHC.

$Antiferromagnetic$ FeSe $monolayer.\quad$ TRS-broken QSH insulators, which are out of $Z_{2}$ classification, are also characterized by the spin Chern number $C_{S}$ \cite{Sheng_PRL2011}. We clarify that, similar to the TRS-preserved cases, the near-quantization of SHC in TRS-broken QSH insulators not only requires a nonzero $C_{S}$ but also necessitates the protection from spin $U(1)$ quasisymmetry. Here we focus on collinear antiferromagnetic configurations where the antiferromagnetic exchange interaction can be described by $m\sigma_{z}\otimes\gamma_{z}$, with $\sigma$ and $\gamma$ denoting the spin and site degrees of freedom, respectively, and $m$ representing the magnitude of exchange splitting. Adding such a collinear antiferromagnetic order to the orbital doublet in Fig. \ref{fig:1}(b), the unperturbed Hamiltonian $H_{0}$ is simply extended by a direct product with $m\gamma_{z}$, thereby retaining the spin $U(1)$ quasisymmetry. Furthermore, using the magnetic point group $6^{\prime}mm^{\prime}$ as an example, we find that the absence of orbital doublets leads to the dominance of spin-mixing interactions \cite{SM}. Therefore, we suggest that orbital doublets are also indispensable in collinear antiferromagnetic scenarios.

\begin{figure}[t]
\includegraphics[width=8.5cm]{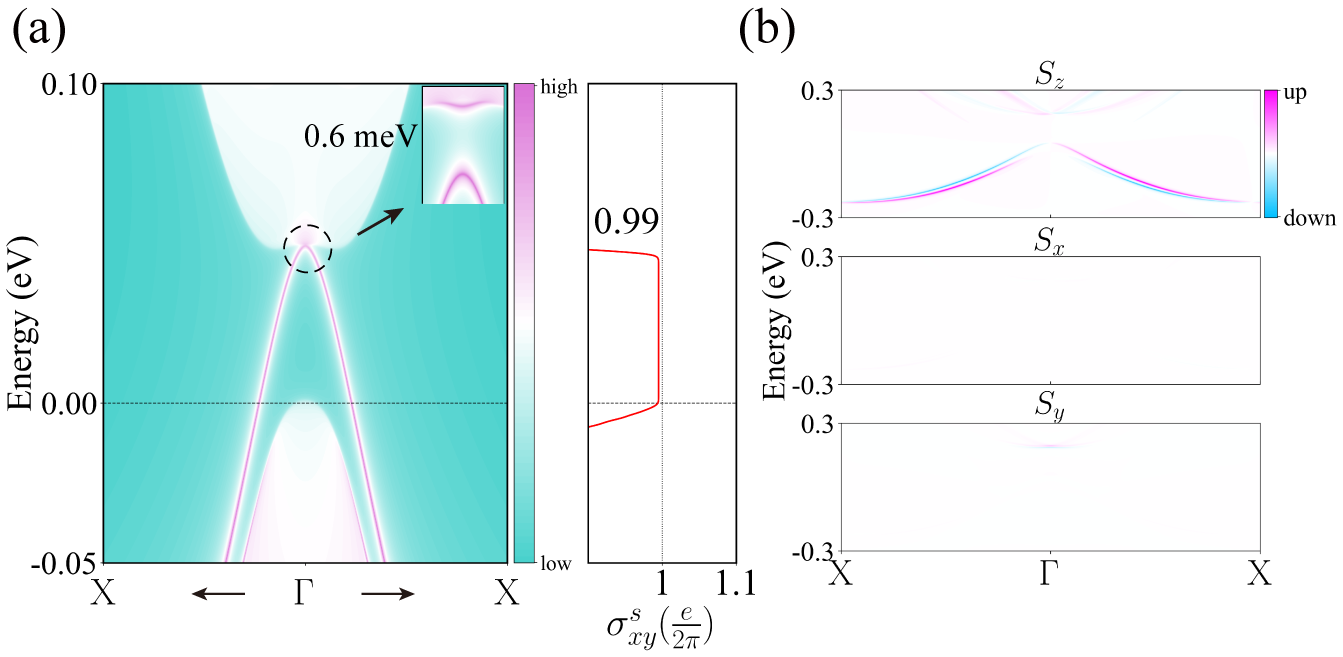}
  \caption{(a) The edge states and SHC, and (b) the spin components of edge states in FeSe monolayer. The bands with sharper purple are more localized on the boundary.}
  \label{fig:4}
\end{figure}

We next perform DFT calculations on antiferromagnetic monolayer FeSe, a recognized TRS-broken QSH insulator \cite{FeSe_PRX,FeSe_NM,FeSe_nl,FeSe_npj}, to validate our theory. In this $C_{S}=1$ system, the low-energy physics is governed by the orbital doublet $\{l_{z+}, l_{z-}\}$ (see Figs. S7 and S8 \cite{SM}). Within such an eigenspace, the dominant SOC Hamiltonian is spin preserved rendering the first-order spin-mixing perturbation $\left\langle \uparrow \right | H^{\prime} \left |\downarrow \right\rangle=0$ [Eq. (\ref{equation2})]. Therefore, the eigenspace of FeSe manifests spin U(1) quasisymmetry. Note that the exchange splitting in FeSe reaches 2-3 eV \cite{FeSe_npj}, which significantly exceeds the SOC strength of approximately 0.3 eV. As shown in Fig. \ref{fig:4}, the inherent spin $U(1)$ quasisymmetry, combined with a significant exchange splitting, leads to a tiny edge gap of 0.6 meV and an almost quantized SHC of 0.99.

$Discussion.\quad$ We clarify that while the exact spin U(1) symmetry does not exist in crystal solids, it can be approximately met accidentally in a case-specific manner. An example is the monolayer WTe$_{2}$, where a persistent spin pattern forms at specific regions of the Brillouin zone \cite{WTe2_prx,WTe2_nl,WTe2_canted}. However, such QSH insulators, dependent on specific chemical environments, can be hardly predicted or designed. Instead, spin U(1) quasisymmetry, serving as a symmetry indicator, establishes a universal framework for predicting and elucidating QSH effect, linking topological invariants and symmetries to the long-sought near-quantized SHC. Notably, it bridges the gap between the widely acknowledged TRS-preserved $Z_{2}=1$ QSH insulators and $Z_{2}=0$ cases, as well as the TRS-broken QSH insulators. While $Z_{2}$ invariant is irrelevant, the two indispensable prerequisites for QSH effect are spin U(1) quasisymmetry and nontrivial $C_{S}$. Overall, our work provides a new perspective for understanding QSH effects and related topological phenomena, and greatly expands the potential material pool for the screening of “best-of-class” material candidates.

%\section*{Acknowledgments}
$Acknowledgments.\quad$ We thank Quansheng Wu, Zhida Song, and Junwei Liu for helpful discussions. Q.L. acknowledges support by National Key R$\&$D Program of China under Grant No. 2020YFA0308900, the National Natural Science Foundation of China under Grant No. 12274194, the Guangdong Provincial Key Laboratory for Computational Science and Material Design under Grant No. 2019B030301001, the Shenzhen Science and Technology Program (No. RCJC20221008092722009 and No.
20231117091158001), the Innovative Team of General Higher Educational Institutes in Guangdong Province (No. 2020KCXTD001), the Science, Technology and Innovation Commission of Shenzhen Municipality under Grant No. ZDSYS20190902092905285, and the Center for Computational Science and Engineering of Southern University of Science and Technology.

%%%REFERENCES%%%
\bibliography{rsc} %You need to replace "rsc" on this line with the name of your .bib file

\begin{thebibliography}{66}%
\makeatletter
\providecommand \@ifxundefined [1]{%
 \@ifx{#1\undefined}
}%
\providecommand \@ifnum [1]{%
 \ifnum #1\expandafter \@firstoftwo
 \else \expandafter \@secondoftwo
 \fi
}%
\providecommand \@ifx [1]{%
 \ifx #1\expandafter \@firstoftwo
 \else \expandafter \@secondoftwo
 \fi
}%
\providecommand \natexlab [1]{#1}%
\providecommand \enquote  [1]{``#1''}%
\providecommand \bibnamefont  [1]{#1}%
\providecommand \bibfnamefont [1]{#1}%
\providecommand \citenamefont [1]{#1}%
\providecommand \href@noop [0]{\@secondoftwo}%
\providecommand \href [0]{\begingroup \@sanitize@url \@href}%
\providecommand \@href[1]{\@@startlink{#1}\@@href}%
\providecommand \@@href[1]{\endgroup#1\@@endlink}%
\providecommand \@sanitize@url [0]{\catcode `\\12\catcode `\$12\catcode
  `\&12\catcode `\#12\catcode `\^12\catcode `\_12\catcode `\%12\relax}%
\providecommand \@@startlink[1]{}%
\providecommand \@@endlink[0]{}%
\providecommand \url  [0]{\begingroup\@sanitize@url \@url }%
\providecommand \@url [1]{\endgroup\@href {#1}{\urlprefix }}%
\providecommand \urlprefix  [0]{URL }%
\providecommand \Eprint [0]{\href }%
\providecommand \doibase [0]{https://doi.org/}%
\providecommand \selectlanguage [0]{\@gobble}%
\providecommand \bibinfo  [0]{\@secondoftwo}%
\providecommand \bibfield  [0]{\@secondoftwo}%
\providecommand \translation [1]{[#1]}%
\providecommand \BibitemOpen [0]{}%
\providecommand \bibitemStop [0]{}%
\providecommand \bibitemNoStop [0]{.\EOS\space}%
\providecommand \EOS [0]{\spacefactor3000\relax}%
\providecommand \BibitemShut  [1]{\csname bibitem#1\endcsname}%
\let\auto@bib@innerbib\@empty
%</preamble>
\bibitem [{\citenamefont {Kane}\ and\ \citenamefont
  {Mele}(2005{\natexlab{a}})}]{QSH_gra}%
  \BibitemOpen
  \bibfield  {author} {\bibinfo {author} {\bibfnamefont {C.~L.}\ \bibnamefont
  {Kane}}\ and\ \bibinfo {author} {\bibfnamefont {E.~J.}\ \bibnamefont
  {Mele}},\ }\bibfield  {title} {\bibinfo {title} {Quantum spin $\mathrm{H}$all
  effect in graphene},\ }\href {https://doi.org/10.1103/PhysRevLett.95.226801}
  {\bibfield  {journal} {\bibinfo  {journal} {Phys. Rev. Lett.}\ }\textbf
  {\bibinfo {volume} {95}},\ \bibinfo {pages} {226801} (\bibinfo {year}
  {2005}{\natexlab{a}})}\BibitemShut {NoStop}%
\bibitem [{\citenamefont {Bernevig}\ and\ \citenamefont
  {Zhang}(2006)}]{QSH_SCZ}%
  \BibitemOpen
  \bibfield  {author} {\bibinfo {author} {\bibfnamefont {B.~A.}\ \bibnamefont
  {Bernevig}}\ and\ \bibinfo {author} {\bibfnamefont {S.-C.}\ \bibnamefont
  {Zhang}},\ }\bibfield  {title} {\bibinfo {title} {Quantum spin {H}all
  effect},\ }\href {https://doi.org/10.1103/PhysRevLett.96.106802} {\bibfield
  {journal} {\bibinfo  {journal} {Phys. Rev. Lett.}\ }\textbf {\bibinfo
  {volume} {96}},\ \bibinfo {pages} {106802} (\bibinfo {year}
  {2006})}\BibitemShut {NoStop}%
\bibitem [{\citenamefont {Bernevig}\ \emph {et~al.}(2006)\citenamefont
  {Bernevig}, \citenamefont {Hughes},\ and\ \citenamefont
  {Zhang}}]{ZSC_Science2006}%
  \BibitemOpen
  \bibfield  {author} {\bibinfo {author} {\bibfnamefont {B.~A.}\ \bibnamefont
  {Bernevig}}, \bibinfo {author} {\bibfnamefont {T.~L.}\ \bibnamefont
  {Hughes}},\ and\ \bibinfo {author} {\bibfnamefont {S.-C.}\ \bibnamefont
  {Zhang}},\ }\bibfield  {title} {\bibinfo {title} {Quantum spin
  $\mathrm{H}$all effect and topological phase transition in
  $\mathrm{H}$g$\mathrm{T}$e quantum wells},\ }\href
  {https://doi.org/10.1126/science.1133734} {\bibfield  {journal} {\bibinfo
  {journal} {Science}\ }\textbf {\bibinfo {volume} {314}},\ \bibinfo {pages}
  {1757} (\bibinfo {year} {2006})}\BibitemShut {NoStop}%
\bibitem [{\citenamefont {K{\"o}nig}\ \emph {et~al.}(2007)\citenamefont
  {K{\"o}nig}, \citenamefont {Wiedmann}, \citenamefont {Br{\"u}ne},
  \citenamefont {Roth}, \citenamefont {Buhmann}, \citenamefont {Molenkamp},
  \citenamefont {Qi},\ and\ \citenamefont {Zhang}}]{HgTe_exp1}%
  \BibitemOpen
  \bibfield  {author} {\bibinfo {author} {\bibfnamefont {M.}~\bibnamefont
  {K{\"o}nig}}, \bibinfo {author} {\bibfnamefont {S.}~\bibnamefont {Wiedmann}},
  \bibinfo {author} {\bibfnamefont {C.}~\bibnamefont {Br{\"u}ne}}, \bibinfo
  {author} {\bibfnamefont {A.}~\bibnamefont {Roth}}, \bibinfo {author}
  {\bibfnamefont {H.}~\bibnamefont {Buhmann}}, \bibinfo {author} {\bibfnamefont
  {L.~W.}\ \bibnamefont {Molenkamp}}, \bibinfo {author} {\bibfnamefont {X.-L.}\
  \bibnamefont {Qi}},\ and\ \bibinfo {author} {\bibfnamefont {S.-C.}\
  \bibnamefont {Zhang}},\ }\bibfield  {title} {\bibinfo {title} {Quantum spin
  $\mathrm{H}$all insulator state in $\mathrm{H}$g$\mathrm{T}$e quantum
  wells},\ }\href {https://doi.org/10.1126/science.1148047} {\bibfield
  {journal} {\bibinfo  {journal} {Science}\ }\textbf {\bibinfo {volume}
  {318}},\ \bibinfo {pages} {766} (\bibinfo {year} {2007})}\BibitemShut
  {NoStop}%
\bibitem [{\citenamefont {Hasan}\ and\ \citenamefont
  {Kane}(2010)}]{TI_RMP2010}%
  \BibitemOpen
  \bibfield  {author} {\bibinfo {author} {\bibfnamefont {M.~Z.}\ \bibnamefont
  {Hasan}}\ and\ \bibinfo {author} {\bibfnamefont {C.~L.}\ \bibnamefont
  {Kane}},\ }\bibfield  {title} {\bibinfo {title} {Colloquium: Topological
  insulators},\ }\href {https://doi.org/10.1103/RevModPhys.82.3045} {\bibfield
  {journal} {\bibinfo  {journal} {Rev. Mod. Phys.}\ }\textbf {\bibinfo {volume}
  {82}},\ \bibinfo {pages} {3045} (\bibinfo {year} {2010})}\BibitemShut
  {NoStop}%
\bibitem [{\citenamefont {Qi}\ and\ \citenamefont {Zhang}(2011)}]{TI_RMP2011}%
  \BibitemOpen
  \bibfield  {author} {\bibinfo {author} {\bibfnamefont {X.-L.}\ \bibnamefont
  {Qi}}\ and\ \bibinfo {author} {\bibfnamefont {S.-C.}\ \bibnamefont {Zhang}},\
  }\bibfield  {title} {\bibinfo {title} {Topological insulators and
  superconductors},\ }\href {https://doi.org/10.1103/RevModPhys.83.1057}
  {\bibfield  {journal} {\bibinfo  {journal} {Rev. Mod. Phys.}\ }\textbf
  {\bibinfo {volume} {83}},\ \bibinfo {pages} {1057} (\bibinfo {year}
  {2011})}\BibitemShut {NoStop}%
\bibitem [{\citenamefont {Knez}\ \emph {et~al.}(2011)\citenamefont {Knez},
  \citenamefont {Du},\ and\ \citenamefont {Sullivan}}]{InAs_exp}%
  \BibitemOpen
  \bibfield  {author} {\bibinfo {author} {\bibfnamefont {I.}~\bibnamefont
  {Knez}}, \bibinfo {author} {\bibfnamefont {R.-R.}\ \bibnamefont {Du}},\ and\
  \bibinfo {author} {\bibfnamefont {G.}~\bibnamefont {Sullivan}},\ }\bibfield
  {title} {\bibinfo {title} {Evidence for helical edge modes in inverted
  $\mathrm{InAs}/\mathrm{GaSb}$ quantum wells},\ }\href
  {https://doi.org/10.1103/PhysRevLett.107.136603} {\bibfield  {journal}
  {\bibinfo  {journal} {Phys. Rev. Lett.}\ }\textbf {\bibinfo {volume} {107}},\
  \bibinfo {pages} {136603} (\bibinfo {year} {2011})}\BibitemShut {NoStop}%
\bibitem [{\citenamefont {Han}\ \emph {et~al.}(2018)\citenamefont {Han},
  \citenamefont {Otani},\ and\ \citenamefont {Maekawa}}]{SOT}%
  \BibitemOpen
  \bibfield  {author} {\bibinfo {author} {\bibfnamefont {W.}~\bibnamefont
  {Han}}, \bibinfo {author} {\bibfnamefont {Y.}~\bibnamefont {Otani}},\ and\
  \bibinfo {author} {\bibfnamefont {S.}~\bibnamefont {Maekawa}},\ }\bibfield
  {title} {\bibinfo {title} {Quantum materials for spin and charge
  conversion},\ }\href {https://doi.org/10.1038/s41535-018-0100-9} {\bibfield
  {journal} {\bibinfo  {journal} {npj Quant. Mater.}\ }\textbf {\bibinfo
  {volume} {3}},\ \bibinfo {pages} {1} (\bibinfo {year} {2018})}\BibitemShut
  {NoStop}%
\bibitem [{\citenamefont {Sajadi}\ \emph {et~al.}(2018)\citenamefont {Sajadi},
  \citenamefont {Palomaki}, \citenamefont {Fei}, \citenamefont {Zhao},
  \citenamefont {Bement}, \citenamefont {Olsen}, \citenamefont {Luescher},
  \citenamefont {Xu}, \citenamefont {Folk},\ and\ \citenamefont
  {Cobden}}]{SC_1}%
  \BibitemOpen
  \bibfield  {author} {\bibinfo {author} {\bibfnamefont {E.}~\bibnamefont
  {Sajadi}}, \bibinfo {author} {\bibfnamefont {T.}~\bibnamefont {Palomaki}},
  \bibinfo {author} {\bibfnamefont {Z.}~\bibnamefont {Fei}}, \bibinfo {author}
  {\bibfnamefont {W.}~\bibnamefont {Zhao}}, \bibinfo {author} {\bibfnamefont
  {P.}~\bibnamefont {Bement}}, \bibinfo {author} {\bibfnamefont
  {C.}~\bibnamefont {Olsen}}, \bibinfo {author} {\bibfnamefont
  {S.}~\bibnamefont {Luescher}}, \bibinfo {author} {\bibfnamefont
  {X.}~\bibnamefont {Xu}}, \bibinfo {author} {\bibfnamefont {J.~A.}\
  \bibnamefont {Folk}},\ and\ \bibinfo {author} {\bibfnamefont {D.~H.}\
  \bibnamefont {Cobden}},\ }\bibfield  {title} {\bibinfo {title} {Gate-induced
  superconductivity in a monolayer topological insulator},\ }\href
  {https://doi.org/10.1126/science.aar4426} {\bibfield  {journal} {\bibinfo
  {journal} {Science}\ }\textbf {\bibinfo {volume} {362}},\ \bibinfo {pages}
  {922} (\bibinfo {year} {2018})}\BibitemShut {NoStop}%
\bibitem [{\citenamefont {Xu}\ \emph {et~al.}(2019)\citenamefont {Xu},
  \citenamefont {Chen}, \citenamefont {Wang}, \citenamefont {Liu},\ and\
  \citenamefont {Ma}}]{transistor}%
  \BibitemOpen
  \bibfield  {author} {\bibinfo {author} {\bibfnamefont {Y.}~\bibnamefont
  {Xu}}, \bibinfo {author} {\bibfnamefont {Y.-R.}\ \bibnamefont {Chen}},
  \bibinfo {author} {\bibfnamefont {J.}~\bibnamefont {Wang}}, \bibinfo {author}
  {\bibfnamefont {J.-F.}\ \bibnamefont {Liu}},\ and\ \bibinfo {author}
  {\bibfnamefont {Z.}~\bibnamefont {Ma}},\ }\bibfield  {title} {\bibinfo
  {title} {Quantized field-effect tunneling between topological edge or
  interface states},\ }\href {https://doi.org/10.1103/PhysRevLett.123.206801}
  {\bibfield  {journal} {\bibinfo  {journal} {Phys. Rev. Lett.}\ }\textbf
  {\bibinfo {volume} {123}},\ \bibinfo {pages} {206801} (\bibinfo {year}
  {2019})}\BibitemShut {NoStop}%
\bibitem [{\citenamefont {Gresta}\ \emph {et~al.}(2019)\citenamefont {Gresta},
  \citenamefont {Real},\ and\ \citenamefont {Arrachea}}]{thermo}%
  \BibitemOpen
  \bibfield  {author} {\bibinfo {author} {\bibfnamefont {D.}~\bibnamefont
  {Gresta}}, \bibinfo {author} {\bibfnamefont {M.}~\bibnamefont {Real}},\ and\
  \bibinfo {author} {\bibfnamefont {L.}~\bibnamefont {Arrachea}},\ }\bibfield
  {title} {\bibinfo {title} {Optimal thermoelectricity with quantum spin {H}all
  edge states},\ }\href {https://doi.org/10.1103/PhysRevLett.123.186801}
  {\bibfield  {journal} {\bibinfo  {journal} {Phys. Rev. Lett.}\ }\textbf
  {\bibinfo {volume} {123}},\ \bibinfo {pages} {186801} (\bibinfo {year}
  {2019})}\BibitemShut {NoStop}%
\bibitem [{\citenamefont {Bhalla}\ \emph {et~al.}(2021)\citenamefont {Bhalla},
  \citenamefont {Deng}, \citenamefont {Wang}, \citenamefont {Wang},\ and\
  \citenamefont {Culcer}}]{nonlinear}%
  \BibitemOpen
  \bibfield  {author} {\bibinfo {author} {\bibfnamefont {P.}~\bibnamefont
  {Bhalla}}, \bibinfo {author} {\bibfnamefont {M.-X.}\ \bibnamefont {Deng}},
  \bibinfo {author} {\bibfnamefont {R.-Q.}\ \bibnamefont {Wang}}, \bibinfo
  {author} {\bibfnamefont {L.}~\bibnamefont {Wang}},\ and\ \bibinfo {author}
  {\bibfnamefont {D.}~\bibnamefont {Culcer}},\ }\bibfield  {title} {\bibinfo
  {title} {Nonlinear ballistic response of quantum spin {H}all edge states},\
  }\href {https://doi.org/10.1103/PhysRevLett.127.206801} {\bibfield  {journal}
  {\bibinfo  {journal} {Phys. Rev. Lett.}\ }\textbf {\bibinfo {volume} {127}},\
  \bibinfo {pages} {206801} (\bibinfo {year} {2021})}\BibitemShut {NoStop}%
\bibitem [{\citenamefont {Kane}\ and\ \citenamefont
  {Mele}(2005{\natexlab{b}})}]{Z2_2005}%
  \BibitemOpen
  \bibfield  {author} {\bibinfo {author} {\bibfnamefont {C.~L.}\ \bibnamefont
  {Kane}}\ and\ \bibinfo {author} {\bibfnamefont {E.~J.}\ \bibnamefont
  {Mele}},\ }\bibfield  {title} {\bibinfo {title} {${Z}_{2}$ topological order
  and the quantum spin $\mathrm{H}$all effect},\ }\href
  {https://doi.org/10.1103/PhysRevLett.95.146802} {\bibfield  {journal}
  {\bibinfo  {journal} {Phys. Rev. Lett.}\ }\textbf {\bibinfo {volume} {95}},\
  \bibinfo {pages} {146802} (\bibinfo {year} {2005}{\natexlab{b}})}\BibitemShut
  {NoStop}%
\bibitem [{\citenamefont {Fei}\ \emph {et~al.}(2017)\citenamefont {Fei},
  \citenamefont {Palomaki}, \citenamefont {Wu}, \citenamefont {Zhao},
  \citenamefont {Cai}, \citenamefont {Sun}, \citenamefont {Nguyen},
  \citenamefont {Finney}, \citenamefont {Xu},\ and\ \citenamefont
  {Cobden}}]{WTe2_NP2}%
  \BibitemOpen
  \bibfield  {author} {\bibinfo {author} {\bibfnamefont {Z.}~\bibnamefont
  {Fei}}, \bibinfo {author} {\bibfnamefont {T.}~\bibnamefont {Palomaki}},
  \bibinfo {author} {\bibfnamefont {S.}~\bibnamefont {Wu}}, \bibinfo {author}
  {\bibfnamefont {W.}~\bibnamefont {Zhao}}, \bibinfo {author} {\bibfnamefont
  {X.}~\bibnamefont {Cai}}, \bibinfo {author} {\bibfnamefont {B.}~\bibnamefont
  {Sun}}, \bibinfo {author} {\bibfnamefont {P.}~\bibnamefont {Nguyen}},
  \bibinfo {author} {\bibfnamefont {J.}~\bibnamefont {Finney}}, \bibinfo
  {author} {\bibfnamefont {X.}~\bibnamefont {Xu}},\ and\ \bibinfo {author}
  {\bibfnamefont {D.~H.}\ \bibnamefont {Cobden}},\ }\bibfield  {title}
  {\bibinfo {title} {Edge conduction in monolayer
  $\mathrm{W}\mathrm{T}$e$_{2}$},\ }\href {https://doi.org/10.1038/nphys4091}
  {\bibfield  {journal} {\bibinfo  {journal} {Nature Phys.}\ }\textbf {\bibinfo
  {volume} {13}},\ \bibinfo {pages} {677} (\bibinfo {year} {2017})}\BibitemShut
  {NoStop}%
\bibitem [{\citenamefont {Wu}\ \emph {et~al.}(2018{\natexlab{a}})\citenamefont
  {Wu}, \citenamefont {Fatemi}, \citenamefont {Gibson}, \citenamefont
  {Watanabe}, \citenamefont {Taniguchi}, \citenamefont {Cava},\ and\
  \citenamefont {Jarillo-Herrero}}]{WTe2_science}%
  \BibitemOpen
  \bibfield  {author} {\bibinfo {author} {\bibfnamefont {S.}~\bibnamefont
  {Wu}}, \bibinfo {author} {\bibfnamefont {V.}~\bibnamefont {Fatemi}}, \bibinfo
  {author} {\bibfnamefont {Q.~D.}\ \bibnamefont {Gibson}}, \bibinfo {author}
  {\bibfnamefont {K.}~\bibnamefont {Watanabe}}, \bibinfo {author}
  {\bibfnamefont {T.}~\bibnamefont {Taniguchi}}, \bibinfo {author}
  {\bibfnamefont {R.~J.}\ \bibnamefont {Cava}},\ and\ \bibinfo {author}
  {\bibfnamefont {P.}~\bibnamefont {Jarillo-Herrero}},\ }\bibfield  {title}
  {\bibinfo {title} {Observation of the quantum spin $\mathrm{H}$all effect up
  to 100 $\mathrm{K}$elvin in a monolayer crystal},\ }\href
  {https://doi.org/10.1126/science.aan6003} {\bibfield  {journal} {\bibinfo
  {journal} {Science}\ }\textbf {\bibinfo {volume} {359}},\ \bibinfo {pages}
  {76} (\bibinfo {year} {2018}{\natexlab{a}})}\BibitemShut {NoStop}%
\bibitem [{\citenamefont {Zhao}\ \emph {et~al.}(2021)\citenamefont {Zhao},
  \citenamefont {Runburg}, \citenamefont {Fei}, \citenamefont {Mutch},
  \citenamefont {Malinowski}, \citenamefont {Sun}, \citenamefont {Huang},
  \citenamefont {Pesin}, \citenamefont {Cui}, \citenamefont {Xu}, \citenamefont
  {Chu},\ and\ \citenamefont {Cobden}}]{WTe2_prx}%
  \BibitemOpen
  \bibfield  {author} {\bibinfo {author} {\bibfnamefont {W.}~\bibnamefont
  {Zhao}}, \bibinfo {author} {\bibfnamefont {E.}~\bibnamefont {Runburg}},
  \bibinfo {author} {\bibfnamefont {Z.}~\bibnamefont {Fei}}, \bibinfo {author}
  {\bibfnamefont {J.}~\bibnamefont {Mutch}}, \bibinfo {author} {\bibfnamefont
  {P.}~\bibnamefont {Malinowski}}, \bibinfo {author} {\bibfnamefont
  {B.}~\bibnamefont {Sun}}, \bibinfo {author} {\bibfnamefont {X.}~\bibnamefont
  {Huang}}, \bibinfo {author} {\bibfnamefont {D.}~\bibnamefont {Pesin}},
  \bibinfo {author} {\bibfnamefont {Y.-T.}\ \bibnamefont {Cui}}, \bibinfo
  {author} {\bibfnamefont {X.}~\bibnamefont {Xu}}, \bibinfo {author}
  {\bibfnamefont {J.-H.}\ \bibnamefont {Chu}},\ and\ \bibinfo {author}
  {\bibfnamefont {D.~H.}\ \bibnamefont {Cobden}},\ }\bibfield  {title}
  {\bibinfo {title} {Determination of the spin axis in quantum spin
  $\mathrm{H}$all insulator candidate monolayer
  $\mathrm{W}\mathrm{T}$e$_{2}$},\ }\href
  {https://doi.org/10.1103/PhysRevX.11.041034} {\bibfield  {journal} {\bibinfo
  {journal} {Phys. Rev. X}\ }\textbf {\bibinfo {volume} {11}},\ \bibinfo
  {pages} {041034} (\bibinfo {year} {2021})}\BibitemShut {NoStop}%
\bibitem [{\citenamefont {Matusalem}\ \emph {et~al.}(2019)\citenamefont
  {Matusalem}, \citenamefont {Marques}, \citenamefont {Teles}, \citenamefont
  {Matthes}, \citenamefont {Furthm\"uller},\ and\ \citenamefont
  {Bechstedt}}]{SHC_prb2019}%
  \BibitemOpen
  \bibfield  {author} {\bibinfo {author} {\bibfnamefont {F.}~\bibnamefont
  {Matusalem}}, \bibinfo {author} {\bibfnamefont {M.}~\bibnamefont {Marques}},
  \bibinfo {author} {\bibfnamefont {L.~K.}\ \bibnamefont {Teles}}, \bibinfo
  {author} {\bibfnamefont {L.}~\bibnamefont {Matthes}}, \bibinfo {author}
  {\bibfnamefont {J.}~\bibnamefont {Furthm\"uller}},\ and\ \bibinfo {author}
  {\bibfnamefont {F.}~\bibnamefont {Bechstedt}},\ }\bibfield  {title} {\bibinfo
  {title} {Quantization of spin $\mathrm{H}$all conductivity in two-dimensional
  topological insulators versus symmetry and spin-orbit interaction},\ }\href
  {https://doi.org/10.1103/PhysRevB.100.245430} {\bibfield  {journal} {\bibinfo
   {journal} {Phys. Rev. B}\ }\textbf {\bibinfo {volume} {100}},\ \bibinfo
  {pages} {245430} (\bibinfo {year} {2019})}\BibitemShut {NoStop}%
\bibitem [{\citenamefont {Garcia}\ \emph {et~al.}(2020)\citenamefont {Garcia},
  \citenamefont {Vila}, \citenamefont {Hsu}, \citenamefont {Waintal},
  \citenamefont {Pereira},\ and\ \citenamefont {Roche}}]{WTe2_canted}%
  \BibitemOpen
  \bibfield  {author} {\bibinfo {author} {\bibfnamefont {J.~H.}\ \bibnamefont
  {Garcia}}, \bibinfo {author} {\bibfnamefont {M.}~\bibnamefont {Vila}},
  \bibinfo {author} {\bibfnamefont {C.-H.}\ \bibnamefont {Hsu}}, \bibinfo
  {author} {\bibfnamefont {X.}~\bibnamefont {Waintal}}, \bibinfo {author}
  {\bibfnamefont {V.~M.}\ \bibnamefont {Pereira}},\ and\ \bibinfo {author}
  {\bibfnamefont {S.}~\bibnamefont {Roche}},\ }\bibfield  {title} {\bibinfo
  {title} {Canted persistent spin texture and quantum spin $\mathrm{H}$all
  effect in {WTe}$_{2}$},\ }\href
  {https://doi.org/10.1103/PhysRevLett.125.256603} {\bibfield  {journal}
  {\bibinfo  {journal} {Phys. Rev. Lett.}\ }\textbf {\bibinfo {volume} {125}},\
  \bibinfo {pages} {256603} (\bibinfo {year} {2020})}\BibitemShut {NoStop}%
\bibitem [{\citenamefont {Tan}\ \emph {et~al.}(2021)\citenamefont {Tan},
  \citenamefont {Deng}, \citenamefont {Zheng}, \citenamefont {Xiang},
  \citenamefont {Albarakati}, \citenamefont {Algarni}, \citenamefont {Farrar},
  \citenamefont {Alzahrani}, \citenamefont {Partridge}, \citenamefont {Yi},
  \citenamefont {Hamilton}, \citenamefont {Wang},\ and\ \citenamefont
  {Wang}}]{WTe2_nl}%
  \BibitemOpen
  \bibfield  {author} {\bibinfo {author} {\bibfnamefont {C.}~\bibnamefont
  {Tan}}, \bibinfo {author} {\bibfnamefont {M.-X.}\ \bibnamefont {Deng}},
  \bibinfo {author} {\bibfnamefont {G.}~\bibnamefont {Zheng}}, \bibinfo
  {author} {\bibfnamefont {F.}~\bibnamefont {Xiang}}, \bibinfo {author}
  {\bibfnamefont {S.}~\bibnamefont {Albarakati}}, \bibinfo {author}
  {\bibfnamefont {M.}~\bibnamefont {Algarni}}, \bibinfo {author} {\bibfnamefont
  {L.}~\bibnamefont {Farrar}}, \bibinfo {author} {\bibfnamefont
  {S.}~\bibnamefont {Alzahrani}}, \bibinfo {author} {\bibfnamefont
  {J.}~\bibnamefont {Partridge}}, \bibinfo {author} {\bibfnamefont {J.~B.}\
  \bibnamefont {Yi}}, \bibinfo {author} {\bibfnamefont {A.~R.}\ \bibnamefont
  {Hamilton}}, \bibinfo {author} {\bibfnamefont {R.-Q.}\ \bibnamefont {Wang}},\
  and\ \bibinfo {author} {\bibfnamefont {L.}~\bibnamefont {Wang}},\ }\bibfield
  {title} {\bibinfo {title} {Spin-momentum locking induced anisotropic
  magnetoresistance in monolayer {WTe}$_{2}$},\ }\href
  {https://doi.org/10.1021/acs.nanolett.1c02329} {\bibfield  {journal}
  {\bibinfo  {journal} {Nano Lett.}\ }\textbf {\bibinfo {volume} {21}},\
  \bibinfo {pages} {9005} (\bibinfo {year} {2021})}\BibitemShut {NoStop}%
\bibitem [{\citenamefont {Tanaka}\ \emph {et~al.}(2011)\citenamefont {Tanaka},
  \citenamefont {Furusaki},\ and\ \citenamefont {Matveev}}]{ex1}%
  \BibitemOpen
  \bibfield  {author} {\bibinfo {author} {\bibfnamefont {Y.}~\bibnamefont
  {Tanaka}}, \bibinfo {author} {\bibfnamefont {A.}~\bibnamefont {Furusaki}},\
  and\ \bibinfo {author} {\bibfnamefont {K.~A.}\ \bibnamefont {Matveev}},\
  }\bibfield  {title} {\bibinfo {title} {Conductance of a helical edge liquid
  coupled to a magnetic impurity},\ }\href
  {https://doi.org/10.1103/PhysRevLett.106.236402} {\bibfield  {journal}
  {\bibinfo  {journal} {Phys. Rev. Lett.}\ }\textbf {\bibinfo {volume} {106}},\
  \bibinfo {pages} {236402} (\bibinfo {year} {2011})}\BibitemShut {NoStop}%
\bibitem [{\citenamefont {Schmidt}\ \emph {et~al.}(2012)\citenamefont
  {Schmidt}, \citenamefont {Rachel}, \citenamefont {von Oppen},\ and\
  \citenamefont {Glazman}}]{ex2}%
  \BibitemOpen
  \bibfield  {author} {\bibinfo {author} {\bibfnamefont {T.~L.}\ \bibnamefont
  {Schmidt}}, \bibinfo {author} {\bibfnamefont {S.}~\bibnamefont {Rachel}},
  \bibinfo {author} {\bibfnamefont {F.}~\bibnamefont {von Oppen}},\ and\
  \bibinfo {author} {\bibfnamefont {L.~I.}\ \bibnamefont {Glazman}},\
  }\bibfield  {title} {\bibinfo {title} {Inelastic electron backscattering in a
  generic helical edge channel},\ }\href
  {https://doi.org/10.1103/PhysRevLett.108.156402} {\bibfield  {journal}
  {\bibinfo  {journal} {Phys. Rev. Lett.}\ }\textbf {\bibinfo {volume} {108}},\
  \bibinfo {pages} {156402} (\bibinfo {year} {2012})}\BibitemShut {NoStop}%
\bibitem [{\citenamefont {Kainaris}\ \emph {et~al.}(2014)\citenamefont
  {Kainaris}, \citenamefont {Gornyi}, \citenamefont {Carr},\ and\ \citenamefont
  {Mirlin}}]{ex3}%
  \BibitemOpen
  \bibfield  {author} {\bibinfo {author} {\bibfnamefont {N.}~\bibnamefont
  {Kainaris}}, \bibinfo {author} {\bibfnamefont {I.~V.}\ \bibnamefont
  {Gornyi}}, \bibinfo {author} {\bibfnamefont {S.~T.}\ \bibnamefont {Carr}},\
  and\ \bibinfo {author} {\bibfnamefont {A.~D.}\ \bibnamefont {Mirlin}},\
  }\bibfield  {title} {\bibinfo {title} {Conductivity of a generic helical
  liquid},\ }\href {https://doi.org/10.1103/PhysRevB.90.075118} {\bibfield
  {journal} {\bibinfo  {journal} {Phys. Rev. B}\ }\textbf {\bibinfo {volume}
  {90}},\ \bibinfo {pages} {075118} (\bibinfo {year} {2014})}\BibitemShut
  {NoStop}%
\bibitem [{\citenamefont {V\"ayrynen}\ \emph {et~al.}(2018)\citenamefont
  {V\"ayrynen}, \citenamefont {Pikulin},\ and\ \citenamefont {Alicea}}]{ex4}%
  \BibitemOpen
  \bibfield  {author} {\bibinfo {author} {\bibfnamefont {J.~I.}\ \bibnamefont
  {V\"ayrynen}}, \bibinfo {author} {\bibfnamefont {D.~I.}\ \bibnamefont
  {Pikulin}},\ and\ \bibinfo {author} {\bibfnamefont {J.}~\bibnamefont
  {Alicea}},\ }\bibfield  {title} {\bibinfo {title} {Noise-induced
  backscattering in a quantum spin {Hall} edge},\ }\href
  {https://doi.org/10.1103/PhysRevLett.121.106601} {\bibfield  {journal}
  {\bibinfo  {journal} {Phys. Rev. Lett.}\ }\textbf {\bibinfo {volume} {121}},\
  \bibinfo {pages} {106601} (\bibinfo {year} {2018})}\BibitemShut {NoStop}%
\bibitem [{\citenamefont {Kang}\ \emph
  {et~al.}(2024{\natexlab{a}})\citenamefont {Kang}, \citenamefont {Qiu},
  \citenamefont {Watanabe}, \citenamefont {Taniguchi}, \citenamefont {Shan},\
  and\ \citenamefont {Mak}}]{twist_arxiv1}%
  \BibitemOpen
  \bibfield  {author} {\bibinfo {author} {\bibfnamefont {K.}~\bibnamefont
  {Kang}}, \bibinfo {author} {\bibfnamefont {Y.}~\bibnamefont {Qiu}}, \bibinfo
  {author} {\bibfnamefont {K.}~\bibnamefont {Watanabe}}, \bibinfo {author}
  {\bibfnamefont {T.}~\bibnamefont {Taniguchi}}, \bibinfo {author}
  {\bibfnamefont {J.}~\bibnamefont {Shan}},\ and\ \bibinfo {author}
  {\bibfnamefont {K.~F.}\ \bibnamefont {Mak}},\ }\bibfield  {title} {\bibinfo
  {title} {Observation of the double quantum spin {H}all phase in moir\'e
  {WS}e$_{2}$},\ }\href {https://arxiv.org/abs/2402.04196} {\bibfield
  {journal} {\bibinfo  {journal} {arXiv: 2402.04196}\ } (\bibinfo {year}
  {2024}{\natexlab{a}})}\BibitemShut {NoStop}%
\bibitem [{\citenamefont {Kang}\ \emph
  {et~al.}(2024{\natexlab{b}})\citenamefont {Kang}, \citenamefont {Shen},
  \citenamefont {Qiu}, \citenamefont {Zeng}, \citenamefont {Xia}, \citenamefont
  {Watanabe}, \citenamefont {Taniguchi}, \citenamefont {Shan},\ and\
  \citenamefont {Mak}}]{twist_arxiv2}%
  \BibitemOpen
  \bibfield  {author} {\bibinfo {author} {\bibfnamefont {K.}~\bibnamefont
  {Kang}}, \bibinfo {author} {\bibfnamefont {B.}~\bibnamefont {Shen}}, \bibinfo
  {author} {\bibfnamefont {Y.}~\bibnamefont {Qiu}}, \bibinfo {author}
  {\bibfnamefont {Y.}~\bibnamefont {Zeng}}, \bibinfo {author} {\bibfnamefont
  {Z.}~\bibnamefont {Xia}}, \bibinfo {author} {\bibfnamefont {K.}~\bibnamefont
  {Watanabe}}, \bibinfo {author} {\bibfnamefont {T.}~\bibnamefont {Taniguchi}},
  \bibinfo {author} {\bibfnamefont {J.}~\bibnamefont {Shan}},\ and\ \bibinfo
  {author} {\bibfnamefont {K.~F.}\ \bibnamefont {Mak}},\ }\bibfield  {title}
  {\bibinfo {title} {Evidence of the fractional quantum spin {Hall} effect in
  moir\'e {MoTe}$_{2}$},\ }\href {https://doi.org/10.1038/s41586-024-07214-5}
  {\bibfield  {journal} {\bibinfo  {journal} {Nature}\ }\textbf {\bibinfo
  {volume} {628}},\ \bibinfo {pages} {522} (\bibinfo {year}
  {2024}{\natexlab{b}})}\BibitemShut {NoStop}%
\bibitem [{\citenamefont {Wang}\ \emph {et~al.}(2016)\citenamefont {Wang},
  \citenamefont {Zhang}, \citenamefont {Liu}, \citenamefont {Liu},
  \citenamefont {Tang}, \citenamefont {Song}, \citenamefont {Zhong},
  \citenamefont {Peng}, \citenamefont {Li}, \citenamefont {Nie}, \citenamefont
  {Wang}, \citenamefont {Zhou}, \citenamefont {Ma}, \citenamefont {Xue},\ and\
  \citenamefont {Liu}}]{FeSe_NM}%
  \BibitemOpen
  \bibfield  {author} {\bibinfo {author} {\bibfnamefont {Z.~F.}\ \bibnamefont
  {Wang}}, \bibinfo {author} {\bibfnamefont {H.}~\bibnamefont {Zhang}},
  \bibinfo {author} {\bibfnamefont {D.}~\bibnamefont {Liu}}, \bibinfo {author}
  {\bibfnamefont {C.}~\bibnamefont {Liu}}, \bibinfo {author} {\bibfnamefont
  {C.}~\bibnamefont {Tang}}, \bibinfo {author} {\bibfnamefont {C.}~\bibnamefont
  {Song}}, \bibinfo {author} {\bibfnamefont {Y.}~\bibnamefont {Zhong}},
  \bibinfo {author} {\bibfnamefont {J.}~\bibnamefont {Peng}}, \bibinfo {author}
  {\bibfnamefont {F.}~\bibnamefont {Li}}, \bibinfo {author} {\bibfnamefont
  {C.}~\bibnamefont {Nie}}, \bibinfo {author} {\bibfnamefont {L.}~\bibnamefont
  {Wang}}, \bibinfo {author} {\bibfnamefont {X.~J.}\ \bibnamefont {Zhou}},
  \bibinfo {author} {\bibfnamefont {X.}~\bibnamefont {Ma}}, \bibinfo {author}
  {\bibfnamefont {Q.~K.}\ \bibnamefont {Xue}},\ and\ \bibinfo {author}
  {\bibfnamefont {F.}~\bibnamefont {Liu}},\ }\bibfield  {title} {\bibinfo
  {title} {Topological edge states in a high-temperature superconductor
  {FeSe}/{SrTiO}$_{3}$(001) film},\ }\href {https://doi.org/10.1038/nmat4686}
  {\bibfield  {journal} {\bibinfo  {journal} {Nature Mater.}\ }\textbf
  {\bibinfo {volume} {15}},\ \bibinfo {pages} {968} (\bibinfo {year}
  {2016})}\BibitemShut {NoStop}%
\bibitem [{\citenamefont {Yuan}\ \emph {et~al.}(2018)\citenamefont {Yuan},
  \citenamefont {Li}, \citenamefont {Liu}, \citenamefont {Deng}, \citenamefont
  {Xu}, \citenamefont {Chen}, \citenamefont {Song}, \citenamefont {Wang},
  \citenamefont {He}, \citenamefont {Xu}, \citenamefont {Ma},\ and\
  \citenamefont {Xue}}]{FeSe_nl}%
  \BibitemOpen
  \bibfield  {author} {\bibinfo {author} {\bibfnamefont {Y.}~\bibnamefont
  {Yuan}}, \bibinfo {author} {\bibfnamefont {W.}~\bibnamefont {Li}}, \bibinfo
  {author} {\bibfnamefont {B.}~\bibnamefont {Liu}}, \bibinfo {author}
  {\bibfnamefont {P.}~\bibnamefont {Deng}}, \bibinfo {author} {\bibfnamefont
  {Z.}~\bibnamefont {Xu}}, \bibinfo {author} {\bibfnamefont {X.}~\bibnamefont
  {Chen}}, \bibinfo {author} {\bibfnamefont {C.}~\bibnamefont {Song}}, \bibinfo
  {author} {\bibfnamefont {L.}~\bibnamefont {Wang}}, \bibinfo {author}
  {\bibfnamefont {K.}~\bibnamefont {He}}, \bibinfo {author} {\bibfnamefont
  {G.}~\bibnamefont {Xu}}, \bibinfo {author} {\bibfnamefont {X.}~\bibnamefont
  {Ma}},\ and\ \bibinfo {author} {\bibfnamefont {Q.-K.}\ \bibnamefont {Xue}},\
  }\bibfield  {title} {\bibinfo {title} {Edge states at nematic domain walls in
  {FeSe} films},\ }\href {https://doi.org/10.1021/acs.nanolett.8b03282}
  {\bibfield  {journal} {\bibinfo  {journal} {Nano Lett.}\ }\textbf {\bibinfo
  {volume} {18}},\ \bibinfo {pages} {7176} (\bibinfo {year}
  {2018})}\BibitemShut {NoStop}%
\bibitem [{\citenamefont {Sheng}\ \emph {et~al.}(2006)\citenamefont {Sheng},
  \citenamefont {Weng}, \citenamefont {Sheng},\ and\ \citenamefont
  {Haldane}}]{Haldane2006}%
  \BibitemOpen
  \bibfield  {author} {\bibinfo {author} {\bibfnamefont {D.~N.}\ \bibnamefont
  {Sheng}}, \bibinfo {author} {\bibfnamefont {Z.~Y.}\ \bibnamefont {Weng}},
  \bibinfo {author} {\bibfnamefont {L.}~\bibnamefont {Sheng}},\ and\ \bibinfo
  {author} {\bibfnamefont {F.~D.~M.}\ \bibnamefont {Haldane}},\ }\bibfield
  {title} {\bibinfo {title} {Quantum spin-$\mathrm{H}$all effect and
  topologically invariant {C}hern numbers},\ }\href
  {https://doi.org/10.1103/PhysRevLett.97.036808} {\bibfield  {journal}
  {\bibinfo  {journal} {Phys. Rev. Lett.}\ }\textbf {\bibinfo {volume} {97}},\
  \bibinfo {pages} {036808} (\bibinfo {year} {2006})}\BibitemShut {NoStop}%
\bibitem [{\citenamefont {Prodan}(2009)}]{SCN_robust}%
  \BibitemOpen
  \bibfield  {author} {\bibinfo {author} {\bibfnamefont {E.}~\bibnamefont
  {Prodan}},\ }\bibfield  {title} {\bibinfo {title} {Robustness of the
  spin-{C}hern number},\ }\href {https://doi.org/10.1103/PhysRevB.80.125327}
  {\bibfield  {journal} {\bibinfo  {journal} {Phys. Rev. B}\ }\textbf {\bibinfo
  {volume} {80}},\ \bibinfo {pages} {125327} (\bibinfo {year}
  {2009})}\BibitemShut {NoStop}%
\bibitem [{\citenamefont {Yang}\ \emph {et~al.}(2011)\citenamefont {Yang},
  \citenamefont {Xu}, \citenamefont {Sheng}, \citenamefont {Wang},
  \citenamefont {Xing},\ and\ \citenamefont {Sheng}}]{Sheng_PRL2011}%
  \BibitemOpen
  \bibfield  {author} {\bibinfo {author} {\bibfnamefont {Y.}~\bibnamefont
  {Yang}}, \bibinfo {author} {\bibfnamefont {Z.}~\bibnamefont {Xu}}, \bibinfo
  {author} {\bibfnamefont {L.}~\bibnamefont {Sheng}}, \bibinfo {author}
  {\bibfnamefont {B.}~\bibnamefont {Wang}}, \bibinfo {author} {\bibfnamefont
  {D.~Y.}\ \bibnamefont {Xing}},\ and\ \bibinfo {author} {\bibfnamefont
  {D.~N.}\ \bibnamefont {Sheng}},\ }\bibfield  {title} {\bibinfo {title}
  {Time-reversal-symmetry-broken quantum spin $\mathrm{H}$all effect},\ }\href
  {https://doi.org/10.1103/PhysRevLett.107.066602} {\bibfield  {journal}
  {\bibinfo  {journal} {Phys. Rev. Lett.}\ }\textbf {\bibinfo {volume} {107}},\
  \bibinfo {pages} {066602} (\bibinfo {year} {2011})}\BibitemShut {NoStop}%
\bibitem [{\citenamefont {Sheng}\ \emph {et~al.}(2005)\citenamefont {Sheng},
  \citenamefont {Sheng}, \citenamefont {Ting},\ and\ \citenamefont
  {Haldane}}]{Haldane2005}%
  \BibitemOpen
  \bibfield  {author} {\bibinfo {author} {\bibfnamefont {L.}~\bibnamefont
  {Sheng}}, \bibinfo {author} {\bibfnamefont {D.~N.}\ \bibnamefont {Sheng}},
  \bibinfo {author} {\bibfnamefont {C.~S.}\ \bibnamefont {Ting}},\ and\
  \bibinfo {author} {\bibfnamefont {F.~D.~M.}\ \bibnamefont {Haldane}},\
  }\bibfield  {title} {\bibinfo {title} {Nondissipative spin $\mathrm{H}$all
  effect via quantized edge transport},\ }\href
  {https://doi.org/10.1103/PhysRevLett.95.136602} {\bibfield  {journal}
  {\bibinfo  {journal} {Phys. Rev. Lett.}\ }\textbf {\bibinfo {volume} {95}},\
  \bibinfo {pages} {136602} (\bibinfo {year} {2005})}\BibitemShut {NoStop}%
\bibitem [{\citenamefont {Wen}(2012)}]{wenxg}%
  \BibitemOpen
  \bibfield  {author} {\bibinfo {author} {\bibfnamefont {X.-G.}\ \bibnamefont
  {Wen}},\ }\bibfield  {title} {\bibinfo {title} {Symmetry-protected
  topological phases in noninteracting fermion systems},\ }\href
  {https://doi.org/10.1103/PhysRevB.85.085103} {\bibfield  {journal} {\bibinfo
  {journal} {Phys. Rev. B}\ }\textbf {\bibinfo {volume} {85}},\ \bibinfo
  {pages} {085103} (\bibinfo {year} {2012})}\BibitemShut {NoStop}%
\bibitem [{\citenamefont {Guo}\ \emph {et~al.}(2022)\citenamefont {Guo},
  \citenamefont {Hu}, \citenamefont {Putzke}, \citenamefont {Diaz},
  \citenamefont {Huang}, \citenamefont {Manna}, \citenamefont {Fan},
  \citenamefont {Shekhar}, \citenamefont {Sun}, \citenamefont {Felser},
  \citenamefont {Liu}, \citenamefont {Bernevig},\ and\ \citenamefont
  {Moll}}]{quasi_np}%
  \BibitemOpen
  \bibfield  {author} {\bibinfo {author} {\bibfnamefont {C.}~\bibnamefont
  {Guo}}, \bibinfo {author} {\bibfnamefont {L.}~\bibnamefont {Hu}}, \bibinfo
  {author} {\bibfnamefont {C.}~\bibnamefont {Putzke}}, \bibinfo {author}
  {\bibfnamefont {J.}~\bibnamefont {Diaz}}, \bibinfo {author} {\bibfnamefont
  {X.}~\bibnamefont {Huang}}, \bibinfo {author} {\bibfnamefont
  {K.}~\bibnamefont {Manna}}, \bibinfo {author} {\bibfnamefont {F.-R.}\
  \bibnamefont {Fan}}, \bibinfo {author} {\bibfnamefont {C.}~\bibnamefont
  {Shekhar}}, \bibinfo {author} {\bibfnamefont {Y.}~\bibnamefont {Sun}},
  \bibinfo {author} {\bibfnamefont {C.}~\bibnamefont {Felser}}, \bibinfo
  {author} {\bibfnamefont {C.}~\bibnamefont {Liu}}, \bibinfo {author}
  {\bibfnamefont {B.~A.}\ \bibnamefont {Bernevig}},\ and\ \bibinfo {author}
  {\bibfnamefont {P.~J.~W.}\ \bibnamefont {Moll}},\ }\bibfield  {title}
  {\bibinfo {title} {Quasi-symmetry-protected topology in a semi-metal},\
  }\href {https://doi.org/10.1038/s41567-022-01604-0} {\bibfield  {journal}
  {\bibinfo  {journal} {Nat. Phys.}\ }\textbf {\bibinfo {volume} {18}},\
  \bibinfo {pages} {813} (\bibinfo {year} {2022})}\BibitemShut {NoStop}%
\bibitem [{\citenamefont {Hu}\ \emph {et~al.}(2023)\citenamefont {Hu},
  \citenamefont {Guo}, \citenamefont {Sun}, \citenamefont {Felser},
  \citenamefont {Elcoro}, \citenamefont {Moll}, \citenamefont {Liu},\ and\
  \citenamefont {Bernevig}}]{quasi_prb}%
  \BibitemOpen
  \bibfield  {author} {\bibinfo {author} {\bibfnamefont {L.-H.}\ \bibnamefont
  {Hu}}, \bibinfo {author} {\bibfnamefont {C.}~\bibnamefont {Guo}}, \bibinfo
  {author} {\bibfnamefont {Y.}~\bibnamefont {Sun}}, \bibinfo {author}
  {\bibfnamefont {C.}~\bibnamefont {Felser}}, \bibinfo {author} {\bibfnamefont
  {L.}~\bibnamefont {Elcoro}}, \bibinfo {author} {\bibfnamefont {P.~J.~W.}\
  \bibnamefont {Moll}}, \bibinfo {author} {\bibfnamefont {C.-X.}\ \bibnamefont
  {Liu}},\ and\ \bibinfo {author} {\bibfnamefont {B.~A.}\ \bibnamefont
  {Bernevig}},\ }\bibfield  {title} {\bibinfo {title} {Hierarchy of
  quasisymmetries and degeneracies in the {CoSi} family of chiral crystal
  materials},\ }\href {https://doi.org/10.1103/PhysRevB.107.125145} {\bibfield
  {journal} {\bibinfo  {journal} {Phys. Rev. B}\ }\textbf {\bibinfo {volume}
  {107}},\ \bibinfo {pages} {125145} (\bibinfo {year} {2023})}\BibitemShut
  {NoStop}%
\bibitem [{\citenamefont {Li}\ \emph {et~al.}(2024)\citenamefont {Li},
  \citenamefont {Zhang}, \citenamefont {Liu},\ and\ \citenamefont
  {Liu}}]{quasi_JY}%
  \BibitemOpen
  \bibfield  {author} {\bibinfo {author} {\bibfnamefont {J.}~\bibnamefont
  {Li}}, \bibinfo {author} {\bibfnamefont {A.}~\bibnamefont {Zhang}}, \bibinfo
  {author} {\bibfnamefont {Y.}~\bibnamefont {Liu}},\ and\ \bibinfo {author}
  {\bibfnamefont {Q.}~\bibnamefont {Liu}},\ }\bibfield  {title} {\bibinfo
  {title} {Group theory on quasisymmetry and protected near degeneracy},\
  }\href {https://doi.org/10.1103/PhysRevLett.133.026402} {\bibfield  {journal}
  {\bibinfo  {journal} {Phys. Rev. Lett.}\ }\textbf {\bibinfo {volume} {133}},\
  \bibinfo {pages} {026402} (\bibinfo {year} {2024})}\BibitemShut {NoStop}%
\bibitem [{SM()}]{SM}%
  \BibitemOpen
  \href@noop {} {}\BibitemShut {NoStop}%
\bibitem [{\citenamefont {Löwdin}(1951)}]{SI1}%
  \BibitemOpen
  \bibfield  {author} {\bibinfo {author} {\bibfnamefont {P.}~\bibnamefont
  {Löwdin}},\ }\bibfield  {title} {\bibinfo {title} {A note on the
  quantum‐mechanical perturbation theory},\ }\href
  {https://doi.org/10.1063/1.1748067} {\bibfield  {journal} {\bibinfo
  {journal} {J. Chem. Phys.}\ }\textbf {\bibinfo {volume} {19}},\ \bibinfo
  {pages} {1396} (\bibinfo {year} {1951})}\BibitemShut {NoStop}%
\bibitem [{\citenamefont {Xiao}\ \emph {et~al.}(2010)\citenamefont {Xiao},
  \citenamefont {Chang},\ and\ \citenamefont {Niu}}]{SI3}%
  \BibitemOpen
  \bibfield  {author} {\bibinfo {author} {\bibfnamefont {D.}~\bibnamefont
  {Xiao}}, \bibinfo {author} {\bibfnamefont {M.-C.}\ \bibnamefont {Chang}},\
  and\ \bibinfo {author} {\bibfnamefont {Q.}~\bibnamefont {Niu}},\ }\bibfield
  {title} {\bibinfo {title} {Berry phase effects on electronic properties},\
  }\href {https://doi.org/10.1103/RevModPhys.82.1959} {\bibfield  {journal}
  {\bibinfo  {journal} {Rev. Mod. Phys.}\ }\textbf {\bibinfo {volume} {82}},\
  \bibinfo {pages} {1959} (\bibinfo {year} {2010})}\BibitemShut {NoStop}%
\bibitem [{\citenamefont {Yao}\ and\ \citenamefont {Fang}(2005)}]{SI4}%
  \BibitemOpen
  \bibfield  {author} {\bibinfo {author} {\bibfnamefont {Y.}~\bibnamefont
  {Yao}}\ and\ \bibinfo {author} {\bibfnamefont {Z.}~\bibnamefont {Fang}},\
  }\bibfield  {title} {\bibinfo {title} {Sign changes of intrinsic spin hall
  effect in semiconductors and simple metals: First-principles calculations},\
  }\href {https://doi.org/10.1103/PhysRevLett.95.156601} {\bibfield  {journal}
  {\bibinfo  {journal} {Phys. Rev. Lett.}\ }\textbf {\bibinfo {volume} {95}},\
  \bibinfo {pages} {156601} (\bibinfo {year} {2005})}\BibitemShut {NoStop}%
\bibitem [{\citenamefont {Blaha}\ \emph {et~al.}(2020)\citenamefont {Blaha},
  \citenamefont {Schwarz}, \citenamefont {Tran}, \citenamefont {Laskowski},
  \citenamefont {Madsen},\ and\ \citenamefont {Marks}}]{SI5}%
  \BibitemOpen
  \bibfield  {author} {\bibinfo {author} {\bibfnamefont {P.}~\bibnamefont
  {Blaha}}, \bibinfo {author} {\bibfnamefont {K.}~\bibnamefont {Schwarz}},
  \bibinfo {author} {\bibfnamefont {F.}~\bibnamefont {Tran}}, \bibinfo {author}
  {\bibfnamefont {R.}~\bibnamefont {Laskowski}}, \bibinfo {author}
  {\bibfnamefont {G.~K.~H.}\ \bibnamefont {Madsen}},\ and\ \bibinfo {author}
  {\bibfnamefont {L.~D.}\ \bibnamefont {Marks}},\ }\bibfield  {title} {\bibinfo
  {title} {{WIEN2k: An APW+lo program for calculating the properties of
  solids}},\ }\href {https://doi.org/10.1063/1.5143061} {\bibfield  {journal}
  {\bibinfo  {journal} {J. Chem. Phys.}\ }\textbf {\bibinfo {volume} {152}},\
  \bibinfo {pages} {074101} (\bibinfo {year} {2020})}\BibitemShut {NoStop}%
\bibitem [{\citenamefont {Imai}\ \emph {et~al.}(2022)\citenamefont {Imai},
  \citenamefont {Nawa}, \citenamefont {Shimizu}, \citenamefont {Yamada},
  \citenamefont {Fujihara}, \citenamefont {Aoyama}, \citenamefont {Takahashi},
  \citenamefont {Okuyama}, \citenamefont {Ohashi}, \citenamefont {Hagihala},
  \citenamefont {Torii}, \citenamefont {Morikawa}, \citenamefont {Terauchi},
  \citenamefont {Kawamata}, \citenamefont {Kato}, \citenamefont {Gotou},
  \citenamefont {Itoh}, \citenamefont {Sato},\ and\ \citenamefont
  {Ohgushi}}]{SI6}%
  \BibitemOpen
  \bibfield  {author} {\bibinfo {author} {\bibfnamefont {Y.}~\bibnamefont
  {Imai}}, \bibinfo {author} {\bibfnamefont {K.}~\bibnamefont {Nawa}}, \bibinfo
  {author} {\bibfnamefont {Y.}~\bibnamefont {Shimizu}}, \bibinfo {author}
  {\bibfnamefont {W.}~\bibnamefont {Yamada}}, \bibinfo {author} {\bibfnamefont
  {H.}~\bibnamefont {Fujihara}}, \bibinfo {author} {\bibfnamefont
  {T.}~\bibnamefont {Aoyama}}, \bibinfo {author} {\bibfnamefont
  {R.}~\bibnamefont {Takahashi}}, \bibinfo {author} {\bibfnamefont
  {D.}~\bibnamefont {Okuyama}}, \bibinfo {author} {\bibfnamefont
  {T.}~\bibnamefont {Ohashi}}, \bibinfo {author} {\bibfnamefont
  {M.}~\bibnamefont {Hagihala}}, \bibinfo {author} {\bibfnamefont
  {S.}~\bibnamefont {Torii}}, \bibinfo {author} {\bibfnamefont
  {D.}~\bibnamefont {Morikawa}}, \bibinfo {author} {\bibfnamefont
  {M.}~\bibnamefont {Terauchi}}, \bibinfo {author} {\bibfnamefont
  {T.}~\bibnamefont {Kawamata}}, \bibinfo {author} {\bibfnamefont
  {M.}~\bibnamefont {Kato}}, \bibinfo {author} {\bibfnamefont {H.}~\bibnamefont
  {Gotou}}, \bibinfo {author} {\bibfnamefont {M.}~\bibnamefont {Itoh}},
  \bibinfo {author} {\bibfnamefont {T.~J.}\ \bibnamefont {Sato}},\ and\
  \bibinfo {author} {\bibfnamefont {K.}~\bibnamefont {Ohgushi}},\ }\bibfield
  {title} {\bibinfo {title} {Zigzag magnetic order in the kitaev spin-liquid
  candidate material {RuBr}$_{3}$ with a honeycomb lattice},\ }\href
  {https://doi.org/10.1103/PhysRevB.105.L041112} {\bibfield  {journal}
  {\bibinfo  {journal} {Phys. Rev. B}\ }\textbf {\bibinfo {volume} {105}},\
  \bibinfo {pages} {L041112} (\bibinfo {year} {2022})}\BibitemShut {NoStop}%
\bibitem [{\citenamefont {Anisimov}\ \emph {et~al.}(1993)\citenamefont
  {Anisimov}, \citenamefont {Solovyev}, \citenamefont {Korotin}, \citenamefont
  {Czy\ifmmode~\dot{z}\else \.{z}\fi{}yk},\ and\ \citenamefont
  {Sawatzky}}]{SI8}%
  \BibitemOpen
  \bibfield  {author} {\bibinfo {author} {\bibfnamefont {V.~I.}\ \bibnamefont
  {Anisimov}}, \bibinfo {author} {\bibfnamefont {I.~V.}\ \bibnamefont
  {Solovyev}}, \bibinfo {author} {\bibfnamefont {M.~A.}\ \bibnamefont
  {Korotin}}, \bibinfo {author} {\bibfnamefont {M.~T.}\ \bibnamefont
  {Czy\ifmmode~\dot{z}\else \.{z}\fi{}yk}},\ and\ \bibinfo {author}
  {\bibfnamefont {G.~A.}\ \bibnamefont {Sawatzky}},\ }\bibfield  {title}
  {\bibinfo {title} {Density-functional theory and {NiO} photoemission
  spectra},\ }\href {https://doi.org/10.1103/PhysRevB.48.16929} {\bibfield
  {journal} {\bibinfo  {journal} {Phys. Rev. B}\ }\textbf {\bibinfo {volume}
  {48}},\ \bibinfo {pages} {16929} (\bibinfo {year} {1993})}\BibitemShut
  {NoStop}%
\bibitem [{\citenamefont {Kaib}\ \emph {et~al.}(2022)\citenamefont {Kaib},
  \citenamefont {Riedl}, \citenamefont {Razpopov}, \citenamefont {Li},
  \citenamefont {Backes}, \citenamefont {Mazin},\ and\ \citenamefont
  {Valent{\'{\i}}}}]{SI9}%
  \BibitemOpen
  \bibfield  {author} {\bibinfo {author} {\bibfnamefont {D.~A.~S.}\
  \bibnamefont {Kaib}}, \bibinfo {author} {\bibfnamefont {K.}~\bibnamefont
  {Riedl}}, \bibinfo {author} {\bibfnamefont {A.}~\bibnamefont {Razpopov}},
  \bibinfo {author} {\bibfnamefont {Y.}~\bibnamefont {Li}}, \bibinfo {author}
  {\bibfnamefont {S.}~\bibnamefont {Backes}}, \bibinfo {author} {\bibfnamefont
  {I.}~\bibnamefont {Mazin}},\ and\ \bibinfo {author} {\bibfnamefont
  {R.}~\bibnamefont {Valent{\'{\i}}}},\ }\bibfield  {title} {\bibinfo {title}
  {Electronic and magnetic properties of the {RuX}$_{3}$ ({X=Cl, Br, I})
  family: two siblings—and a cousin?},\ }\href
  {https://doi.org/10.1038/s41535-022-00481-3} {\bibfield  {journal} {\bibinfo
  {journal} {npj Quantum Mater.}\ }\textbf {\bibinfo {volume} {7}},\ \bibinfo
  {pages} {75} (\bibinfo {year} {2022})}\BibitemShut {NoStop}%
\bibitem [{\citenamefont {Zhang}\ \emph {et~al.}(2022)\citenamefont {Zhang},
  \citenamefont {Lin}, \citenamefont {Moreo},\ and\ \citenamefont
  {Dagotto}}]{SI10}%
  \BibitemOpen
  \bibfield  {author} {\bibinfo {author} {\bibfnamefont {Y.}~\bibnamefont
  {Zhang}}, \bibinfo {author} {\bibfnamefont {L.-F.}\ \bibnamefont {Lin}},
  \bibinfo {author} {\bibfnamefont {A.}~\bibnamefont {Moreo}},\ and\ \bibinfo
  {author} {\bibfnamefont {E.}~\bibnamefont {Dagotto}},\ }\bibfield  {title}
  {\bibinfo {title} {Theoretical study of the crystal and electronic properties
  of $\alpha$-{RuI}$_{3}$},\ }\href
  {https://doi.org/10.1103/PhysRevB.105.085107} {\bibfield  {journal} {\bibinfo
   {journal} {Phys. Rev. B}\ }\textbf {\bibinfo {volume} {105}},\ \bibinfo
  {pages} {085107} (\bibinfo {year} {2022})}\BibitemShut {NoStop}%
\bibitem [{\citenamefont {Liu}\ \emph {et~al.}(2023)\citenamefont {Liu},
  \citenamefont {Yang}, \citenamefont {Wang}, \citenamefont {Lu}, \citenamefont
  {Ma},\ and\ \citenamefont {Wu}}]{SI11}%
  \BibitemOpen
  \bibfield  {author} {\bibinfo {author} {\bibfnamefont {L.}~\bibnamefont
  {Liu}}, \bibinfo {author} {\bibfnamefont {K.}~\bibnamefont {Yang}}, \bibinfo
  {author} {\bibfnamefont {G.}~\bibnamefont {Wang}}, \bibinfo {author}
  {\bibfnamefont {D.}~\bibnamefont {Lu}}, \bibinfo {author} {\bibfnamefont
  {Y.}~\bibnamefont {Ma}},\ and\ \bibinfo {author} {\bibfnamefont
  {H.}~\bibnamefont {Wu}},\ }\bibfield  {title} {\bibinfo {title} {Contrasting
  electronic states of {RuI}$_{3}$ and {RuCl}$_{3}$},\ }\href
  {https://doi.org/10.1103/PhysRevB.107.165134} {\bibfield  {journal} {\bibinfo
   {journal} {Phys. Rev. B}\ }\textbf {\bibinfo {volume} {107}},\ \bibinfo
  {pages} {165134} (\bibinfo {year} {2023})}\BibitemShut {NoStop}%
\bibitem [{\citenamefont {Dai}\ and\ \citenamefont {Whangbo}(2005)}]{SI13}%
  \BibitemOpen
  \bibfield  {author} {\bibinfo {author} {\bibfnamefont {D.}~\bibnamefont
  {Dai}}\ and\ \bibinfo {author} {\bibfnamefont {M.-H.}\ \bibnamefont
  {Whangbo}},\ }\bibfield  {title} {\bibinfo {title} {Analysis of the uniaxial
  magnetic properties of high-spin d$^{6}$ ions at trigonal prism and linear
  two-coordinate sites: Uniaxial magnetic properties of
  {Ca}$_{3}${Co}$_{2}${O}$_{6}$ and {Fe}[{C(SiMe}$_{3}$)$_{3}$]$_{2}$},\ }\href
  {https://doi.org/10.1021/ic050185g} {\bibfield  {journal} {\bibinfo
  {journal} {Inorg. Chem.}\ }\textbf {\bibinfo {volume} {44}},\ \bibinfo
  {pages} {4407} (\bibinfo {year} {2005})}\BibitemShut {NoStop}%
\bibitem [{\citenamefont {Kuneš}\ \emph {et~al.}(2010)\citenamefont {Kuneš},
  \citenamefont {Arita}, \citenamefont {Wissgott}, \citenamefont {Toschi},
  \citenamefont {Ikeda},\ and\ \citenamefont {Held}}]{SI14}%
  \BibitemOpen
  \bibfield  {author} {\bibinfo {author} {\bibfnamefont {J.}~\bibnamefont
  {Kuneš}}, \bibinfo {author} {\bibfnamefont {R.}~\bibnamefont {Arita}},
  \bibinfo {author} {\bibfnamefont {P.}~\bibnamefont {Wissgott}}, \bibinfo
  {author} {\bibfnamefont {A.}~\bibnamefont {Toschi}}, \bibinfo {author}
  {\bibfnamefont {H.}~\bibnamefont {Ikeda}},\ and\ \bibinfo {author}
  {\bibfnamefont {K.}~\bibnamefont {Held}},\ }\bibfield  {title} {\bibinfo
  {title} {Wien2wannier: From linearized augmented plane waves to maximally
  localized wannier functions},\ }\href
  {https://doi.org/https://doi.org/10.1016/j.cpc.2010.08.005} {\bibfield
  {journal} {\bibinfo  {journal} {Comp. Phys. Commun.}\ }\textbf {\bibinfo
  {volume} {181}},\ \bibinfo {pages} {1888} (\bibinfo {year}
  {2010})}\BibitemShut {NoStop}%
\bibitem [{\citenamefont {Mostofi}\ \emph {et~al.}(2008)\citenamefont
  {Mostofi}, \citenamefont {Yates}, \citenamefont {Lee}, \citenamefont {Souza},
  \citenamefont {Vanderbilt},\ and\ \citenamefont {Marzari}}]{SI15}%
  \BibitemOpen
  \bibfield  {author} {\bibinfo {author} {\bibfnamefont {A.~A.}\ \bibnamefont
  {Mostofi}}, \bibinfo {author} {\bibfnamefont {J.~R.}\ \bibnamefont {Yates}},
  \bibinfo {author} {\bibfnamefont {Y.-S.}\ \bibnamefont {Lee}}, \bibinfo
  {author} {\bibfnamefont {I.}~\bibnamefont {Souza}}, \bibinfo {author}
  {\bibfnamefont {D.}~\bibnamefont {Vanderbilt}},\ and\ \bibinfo {author}
  {\bibfnamefont {N.}~\bibnamefont {Marzari}},\ }\bibfield  {title} {\bibinfo
  {title} {wannier90: A tool for obtaining maximally-localised wannier
  functions},\ }\href
  {https://doi.org/https://doi.org/10.1016/j.cpc.2007.11.016} {\bibfield
  {journal} {\bibinfo  {journal} {Comput. Phys. Commun.}\ }\textbf {\bibinfo
  {volume} {178}},\ \bibinfo {pages} {685} (\bibinfo {year}
  {2008})}\BibitemShut {NoStop}%
\bibitem [{\citenamefont {Wu}\ \emph {et~al.}(2018{\natexlab{b}})\citenamefont
  {Wu}, \citenamefont {Zhang}, \citenamefont {Song}, \citenamefont {Troyer},\
  and\ \citenamefont {Soluyanov}}]{SI16}%
  \BibitemOpen
  \bibfield  {author} {\bibinfo {author} {\bibfnamefont {Q.}~\bibnamefont
  {Wu}}, \bibinfo {author} {\bibfnamefont {S.}~\bibnamefont {Zhang}}, \bibinfo
  {author} {\bibfnamefont {H.-F.}\ \bibnamefont {Song}}, \bibinfo {author}
  {\bibfnamefont {M.}~\bibnamefont {Troyer}},\ and\ \bibinfo {author}
  {\bibfnamefont {A.~A.}\ \bibnamefont {Soluyanov}},\ }\bibfield  {title}
  {\bibinfo {title} {Wanniertools: An open-source software package for novel
  topological materials},\ }\href
  {https://doi.org/https://doi.org/10.1016/j.cpc.2017.09.033} {\bibfield
  {journal} {\bibinfo  {journal} {Comput. Phys. Commun.}\ }\textbf {\bibinfo
  {volume} {224}},\ \bibinfo {pages} {405} (\bibinfo {year}
  {2018}{\natexlab{b}})}\BibitemShut {NoStop}%
\bibitem [{\citenamefont {Deng}\ \emph {et~al.}(2022)\citenamefont {Deng},
  \citenamefont {Shao}, \citenamefont {Gao}, \citenamefont {Yue}, \citenamefont
  {Weng}, \citenamefont {Fang},\ and\ \citenamefont {Wang}}]{SI21}%
  \BibitemOpen
  \bibfield  {author} {\bibinfo {author} {\bibfnamefont {J.}~\bibnamefont
  {Deng}}, \bibinfo {author} {\bibfnamefont {D.}~\bibnamefont {Shao}}, \bibinfo
  {author} {\bibfnamefont {J.}~\bibnamefont {Gao}}, \bibinfo {author}
  {\bibfnamefont {C.}~\bibnamefont {Yue}}, \bibinfo {author} {\bibfnamefont
  {H.}~\bibnamefont {Weng}}, \bibinfo {author} {\bibfnamefont {Z.}~\bibnamefont
  {Fang}},\ and\ \bibinfo {author} {\bibfnamefont {Z.}~\bibnamefont {Wang}},\
  }\bibfield  {title} {\bibinfo {title} {Twisted nodal wires and
  three-dimensional quantum spin hall effect in distorted square-net
  compounds},\ }\href {https://doi.org/10.1103/PhysRevB.105.224103} {\bibfield
  {journal} {\bibinfo  {journal} {Phys. Rev. B}\ }\textbf {\bibinfo {volume}
  {105}},\ \bibinfo {pages} {224103} (\bibinfo {year} {2022})}\BibitemShut
  {NoStop}%
\bibitem [{\citenamefont {Lin}\ \emph {et~al.}(2024)\citenamefont {Lin},
  \citenamefont {Palumbo}, \citenamefont {Guo}, \citenamefont {Hwang},
  \citenamefont {Blackburn}, \citenamefont {Shoemaker}, \citenamefont
  {Mahmood}, \citenamefont {Wang}, \citenamefont {Fiete}, \citenamefont
  {Wieder},\ and\ \citenamefont {Bradlyn}}]{SI22}%
  \BibitemOpen
  \bibfield  {author} {\bibinfo {author} {\bibfnamefont {K.-S.}\ \bibnamefont
  {Lin}}, \bibinfo {author} {\bibfnamefont {G.}~\bibnamefont {Palumbo}},
  \bibinfo {author} {\bibfnamefont {Z.}~\bibnamefont {Guo}}, \bibinfo {author}
  {\bibfnamefont {Y.}~\bibnamefont {Hwang}}, \bibinfo {author} {\bibfnamefont
  {J.}~\bibnamefont {Blackburn}}, \bibinfo {author} {\bibfnamefont {D.~P.}\
  \bibnamefont {Shoemaker}}, \bibinfo {author} {\bibfnamefont {F.}~\bibnamefont
  {Mahmood}}, \bibinfo {author} {\bibfnamefont {Z.}~\bibnamefont {Wang}},
  \bibinfo {author} {\bibfnamefont {G.~A.}\ \bibnamefont {Fiete}}, \bibinfo
  {author} {\bibfnamefont {B.~J.}\ \bibnamefont {Wieder}},\ and\ \bibinfo
  {author} {\bibfnamefont {B.}~\bibnamefont {Bradlyn}},\ }\bibfield  {title}
  {\bibinfo {title} {Spin-resolved topology and partial axion angles in
  three-dimensional insulators},\ }\href
  {https://doi.org/10.1038/s41467-024-44762-w} {\bibfield  {journal} {\bibinfo
  {journal} {Nature Commun.}\ }\textbf {\bibinfo {volume} {15}},\ \bibinfo
  {pages} {550} (\bibinfo {year} {2024})}\BibitemShut {NoStop}%
\bibitem [{\citenamefont {Gonz\'alez-Hern\'andez}\ and\ \citenamefont
  {Uribe}(2024)}]{SI23}%
  \BibitemOpen
  \bibfield  {author} {\bibinfo {author} {\bibfnamefont {R.}~\bibnamefont
  {Gonz\'alez-Hern\'andez}}\ and\ \bibinfo {author} {\bibfnamefont
  {B.}~\bibnamefont {Uribe}},\ }\bibfield  {title} {\bibinfo {title} {Average
  spin chern number},\ }\href {https://doi.org/10.1103/PhysRevB.110.125129}
  {\bibfield  {journal} {\bibinfo  {journal} {Phys. Rev. B}\ }\textbf {\bibinfo
  {volume} {110}},\ \bibinfo {pages} {125129} (\bibinfo {year}
  {2024})}\BibitemShut {NoStop}%
\bibitem [{\citenamefont {Yao}\ \emph {et~al.}(2024)\citenamefont {Yao},
  \citenamefont {Zhou}, \citenamefont {Hung}, \citenamefont {Lin},
  \citenamefont {Bansil},\ and\ \citenamefont {Chang}}]{SI24}%
  \BibitemOpen
  \bibfield  {author} {\bibinfo {author} {\bibfnamefont {Y.-T.}\ \bibnamefont
  {Yao}}, \bibinfo {author} {\bibfnamefont {X.}~\bibnamefont {Zhou}}, \bibinfo
  {author} {\bibfnamefont {Y.-C.}\ \bibnamefont {Hung}}, \bibinfo {author}
  {\bibfnamefont {H.}~\bibnamefont {Lin}}, \bibinfo {author} {\bibfnamefont
  {A.}~\bibnamefont {Bansil}},\ and\ \bibinfo {author} {\bibfnamefont {T.-R.}\
  \bibnamefont {Chang}},\ }\bibfield  {title} {\bibinfo {title} {Feature-energy
  duality of topological boundary states in a multilayer quantum spin hall
  insulator},\ }\href {https://doi.org/10.1103/PhysRevB.109.155143} {\bibfield
  {journal} {\bibinfo  {journal} {Phys. Rev. B}\ }\textbf {\bibinfo {volume}
  {109}},\ \bibinfo {pages} {155143} (\bibinfo {year} {2024})}\BibitemShut
  {NoStop}%
\bibitem [{\citenamefont {Peng}\ \emph {et~al.}(2024)\citenamefont {Peng},
  \citenamefont {Lange}, \citenamefont {Bennett}, \citenamefont {Wang},
  \citenamefont {Slager},\ and\ \citenamefont {Monserrat}}]{SI25}%
  \BibitemOpen
  \bibfield  {author} {\bibinfo {author} {\bibfnamefont {B.}~\bibnamefont
  {Peng}}, \bibinfo {author} {\bibfnamefont {G.~F.}\ \bibnamefont {Lange}},
  \bibinfo {author} {\bibfnamefont {D.}~\bibnamefont {Bennett}}, \bibinfo
  {author} {\bibfnamefont {K.}~\bibnamefont {Wang}}, \bibinfo {author}
  {\bibfnamefont {R.-J.}\ \bibnamefont {Slager}},\ and\ \bibinfo {author}
  {\bibfnamefont {B.}~\bibnamefont {Monserrat}},\ }\bibfield  {title} {\bibinfo
  {title} {Photoinduced electronic and spin topological phase transitions in
  monolayer bismuth},\ }\href {https://doi.org/10.1103/PhysRevLett.132.116601}
  {\bibfield  {journal} {\bibinfo  {journal} {Phys. Rev. Lett.}\ }\textbf
  {\bibinfo {volume} {132}},\ \bibinfo {pages} {116601} (\bibinfo {year}
  {2024})}\BibitemShut {NoStop}%
\bibitem [{\citenamefont {Tyner}(2024)}]{SI26}%
  \BibitemOpen
  \bibfield  {author} {\bibinfo {author} {\bibfnamefont {A.~C.}\ \bibnamefont
  {Tyner}},\ }\bibfield  {title} {\bibinfo {title} {Machine learning guided
  discovery of stable, spin-resolved topological insulators},\ }\href
  {https://doi.org/10.1103/PhysRevResearch.6.023316} {\bibfield  {journal}
  {\bibinfo  {journal} {Phys. Rev. Res.}\ }\textbf {\bibinfo {volume} {6}},\
  \bibinfo {pages} {023316} (\bibinfo {year} {2024})}\BibitemShut {NoStop}%
\bibitem [{\citenamefont {Yao}\ \emph {et~al.}(2004)\citenamefont {Yao},
  \citenamefont {Kleinman}, \citenamefont {MacDonald}, \citenamefont {Sinova},
  \citenamefont {Jungwirth}, \citenamefont {Wang}, \citenamefont {Wang},\ and\
  \citenamefont {Niu}}]{SI28}%
  \BibitemOpen
  \bibfield  {author} {\bibinfo {author} {\bibfnamefont {Y.}~\bibnamefont
  {Yao}}, \bibinfo {author} {\bibfnamefont {L.}~\bibnamefont {Kleinman}},
  \bibinfo {author} {\bibfnamefont {A.~H.}\ \bibnamefont {MacDonald}}, \bibinfo
  {author} {\bibfnamefont {J.}~\bibnamefont {Sinova}}, \bibinfo {author}
  {\bibfnamefont {T.}~\bibnamefont {Jungwirth}}, \bibinfo {author}
  {\bibfnamefont {D.-s.}\ \bibnamefont {Wang}}, \bibinfo {author}
  {\bibfnamefont {E.}~\bibnamefont {Wang}},\ and\ \bibinfo {author}
  {\bibfnamefont {Q.}~\bibnamefont {Niu}},\ }\bibfield  {title} {\bibinfo
  {title} {First principles calculation of anomalous $\mathrm{H}$all
  conductivity in ferromagnetic bcc {Fe}},\ }\href
  {https://doi.org/10.1103/PhysRevLett.92.037204} {\bibfield  {journal}
  {\bibinfo  {journal} {Phys. Rev. Lett.}\ }\textbf {\bibinfo {volume} {92}},\
  \bibinfo {pages} {037204} (\bibinfo {year} {2004})}\BibitemShut {NoStop}%
\bibitem [{\citenamefont {Xiao}\ \emph {et~al.}(2012)\citenamefont {Xiao},
  \citenamefont {Liu}, \citenamefont {Feng}, \citenamefont {Xu},\ and\
  \citenamefont {Yao}}]{valley_yao}%
  \BibitemOpen
  \bibfield  {author} {\bibinfo {author} {\bibfnamefont {D.}~\bibnamefont
  {Xiao}}, \bibinfo {author} {\bibfnamefont {G.-B.}\ \bibnamefont {Liu}},
  \bibinfo {author} {\bibfnamefont {W.}~\bibnamefont {Feng}}, \bibinfo {author}
  {\bibfnamefont {X.}~\bibnamefont {Xu}},\ and\ \bibinfo {author}
  {\bibfnamefont {W.}~\bibnamefont {Yao}},\ }\bibfield  {title} {\bibinfo
  {title} {Coupled spin and valley physics in monolayers of {M}o{S}$_{2}$ and
  other group-{VI} dichalcogenides},\ }\href
  {https://doi.org/10.1103/PhysRevLett.108.196802} {\bibfield  {journal}
  {\bibinfo  {journal} {Phys. Rev. Lett.}\ }\textbf {\bibinfo {volume} {108}},\
  \bibinfo {pages} {196802} (\bibinfo {year} {2012})}\BibitemShut {NoStop}%
\bibitem [{\citenamefont {Zhang}\ \emph {et~al.}(2024)\citenamefont {Zhang},
  \citenamefont {Chen}, \citenamefont {Li}, \citenamefont {Liu}, \citenamefont
  {Liu},\ and\ \citenamefont {Liu}}]{aozhang}%
  \BibitemOpen
  \bibfield  {author} {\bibinfo {author} {\bibfnamefont {A.}~\bibnamefont
  {Zhang}}, \bibinfo {author} {\bibfnamefont {X.}~\bibnamefont {Chen}},
  \bibinfo {author} {\bibfnamefont {J.}~\bibnamefont {Li}}, \bibinfo {author}
  {\bibfnamefont {P.}~\bibnamefont {Liu}}, \bibinfo {author} {\bibfnamefont
  {Y.}~\bibnamefont {Liu}},\ and\ \bibinfo {author} {\bibfnamefont
  {Q.}~\bibnamefont {Liu}},\ }\bibfield  {title} {\bibinfo {title} {Topological
  charge quadrupole protected by spin-orbit ${U}$(1) quasi-symmetry in
  antiferromagnet {NdBiPt}},\ }\href {https://arxiv.org/abs/2408.07887}
  {\bibfield  {journal} {\bibinfo  {journal} {arXiv: 2408.07887}\ } (\bibinfo
  {year} {2024})}\BibitemShut {NoStop}%
\bibitem [{\citenamefont {Costa}\ \emph {et~al.}(2023)\citenamefont {Costa},
  \citenamefont {Focassio}, \citenamefont {Canonico}, \citenamefont {Cysne},
  \citenamefont {Schleder}, \citenamefont {Muniz}, \citenamefont {Fazzio},\
  and\ \citenamefont {Rappoport}}]{OHE}%
  \BibitemOpen
  \bibfield  {author} {\bibinfo {author} {\bibfnamefont {M.}~\bibnamefont
  {Costa}}, \bibinfo {author} {\bibfnamefont {B.}~\bibnamefont {Focassio}},
  \bibinfo {author} {\bibfnamefont {L.~M.}\ \bibnamefont {Canonico}}, \bibinfo
  {author} {\bibfnamefont {T.~P.}\ \bibnamefont {Cysne}}, \bibinfo {author}
  {\bibfnamefont {G.~R.}\ \bibnamefont {Schleder}}, \bibinfo {author}
  {\bibfnamefont {R.~B.}\ \bibnamefont {Muniz}}, \bibinfo {author}
  {\bibfnamefont {A.}~\bibnamefont {Fazzio}},\ and\ \bibinfo {author}
  {\bibfnamefont {T.~G.}\ \bibnamefont {Rappoport}},\ }\bibfield  {title}
  {\bibinfo {title} {Connecting higher-order topology with the orbital {H}all
  effect in monolayers of transition metal dichalcogenides},\ }\href
  {https://doi.org/10.1103/PhysRevLett.130.116204} {\bibfield  {journal}
  {\bibinfo  {journal} {Phys. Rev. Lett.}\ }\textbf {\bibinfo {volume} {130}},\
  \bibinfo {pages} {116204} (\bibinfo {year} {2023})}\BibitemShut {NoStop}%
\bibitem [{\citenamefont {Wu}\ \emph {et~al.}(2019)\citenamefont {Wu},
  \citenamefont {Lovorn}, \citenamefont {Tutuc}, \citenamefont {Martin},\ and\
  \citenamefont {MacDonald}}]{twist_MCD}%
  \BibitemOpen
  \bibfield  {author} {\bibinfo {author} {\bibfnamefont {F.}~\bibnamefont
  {Wu}}, \bibinfo {author} {\bibfnamefont {T.}~\bibnamefont {Lovorn}}, \bibinfo
  {author} {\bibfnamefont {E.}~\bibnamefont {Tutuc}}, \bibinfo {author}
  {\bibfnamefont {I.}~\bibnamefont {Martin}},\ and\ \bibinfo {author}
  {\bibfnamefont {A.~H.}\ \bibnamefont {MacDonald}},\ }\bibfield  {title}
  {\bibinfo {title} {Topological insulators in twisted transition metal
  dichalcogenide homobilayers},\ }\href
  {https://doi.org/10.1103/PhysRevLett.122.086402} {\bibfield  {journal}
  {\bibinfo  {journal} {Phys. Rev. Lett.}\ }\textbf {\bibinfo {volume} {122}},\
  \bibinfo {pages} {086402} (\bibinfo {year} {2019})}\BibitemShut {NoStop}%
\bibitem [{\citenamefont {Devakul}\ \emph {et~al.}(2021)\citenamefont
  {Devakul}, \citenamefont {Crépel}, \citenamefont {Zhang},\ and\
  \citenamefont {Fu}}]{twist_fu}%
  \BibitemOpen
  \bibfield  {author} {\bibinfo {author} {\bibfnamefont {T.}~\bibnamefont
  {Devakul}}, \bibinfo {author} {\bibfnamefont {V.}~\bibnamefont {Crépel}},
  \bibinfo {author} {\bibfnamefont {Y.}~\bibnamefont {Zhang}},\ and\ \bibinfo
  {author} {\bibfnamefont {L.}~\bibnamefont {Fu}},\ }\bibfield  {title}
  {\bibinfo {title} {Magic in twisted transition metal dichalcogenide
  bilayers},\ }\href {https://doi.org/10.1038/s41467-021-27042-9} {\bibfield
  {journal} {\bibinfo  {journal} {Nature Commun.}\ }\textbf {\bibinfo {volume}
  {12}},\ \bibinfo {pages} {6730} (\bibinfo {year} {2021})}\BibitemShut
  {NoStop}%
\bibitem [{\citenamefont {Liu}\ \emph {et~al.}(2024)\citenamefont {Liu},
  \citenamefont {Liu}, \citenamefont {Li}, \citenamefont {Wu},\ and\
  \citenamefont {Liu}}]{lliu}%
  \BibitemOpen
  \bibfield  {author} {\bibinfo {author} {\bibfnamefont {L.}~\bibnamefont
  {Liu}}, \bibinfo {author} {\bibfnamefont {Y.}~\bibnamefont {Liu}}, \bibinfo
  {author} {\bibfnamefont {J.}~\bibnamefont {Li}}, \bibinfo {author}
  {\bibfnamefont {H.}~\bibnamefont {Wu}},\ and\ \bibinfo {author}
  {\bibfnamefont {Q.}~\bibnamefont {Liu}},\ }\bibfield  {title} {\bibinfo
  {title} {Orbital doublet driven even-spin chern insulators},\ }\href
  {https://doi.org/10.1103/PhysRevB.110.035161} {\bibfield  {journal} {\bibinfo
   {journal} {Phys. Rev. B}\ }\textbf {\bibinfo {volume} {110}},\ \bibinfo
  {pages} {035161} (\bibinfo {year} {2024})}\BibitemShut {NoStop}%
\bibitem [{\citenamefont {Fu}\ and\ \citenamefont {Kane}(2007)}]{FuKane_Z2}%
  \BibitemOpen
  \bibfield  {author} {\bibinfo {author} {\bibfnamefont {L.}~\bibnamefont
  {Fu}}\ and\ \bibinfo {author} {\bibfnamefont {C.~L.}\ \bibnamefont {Kane}},\
  }\bibfield  {title} {\bibinfo {title} {Topological insulators with inversion
  symmetry},\ }\href {https://doi.org/10.1103/PhysRevB.76.045302} {\bibfield
  {journal} {\bibinfo  {journal} {Phys. Rev. B}\ }\textbf {\bibinfo {volume}
  {76}},\ \bibinfo {pages} {045302} (\bibinfo {year} {2007})}\BibitemShut
  {NoStop}%
\bibitem [{\citenamefont {Lv}\ \emph {et~al.}(2021)\citenamefont {Lv},
  \citenamefont {Du}, \citenamefont {Liang}, \citenamefont {Liu}, \citenamefont
  {Liang}, \citenamefont {Chen}, \citenamefont {Zhou}, \citenamefont {Zhang},
  \citenamefont {Zhang}, \citenamefont {Ai}, \citenamefont {Yan},\ and\
  \citenamefont {Zhu}}]{SCN_exp}%
  \BibitemOpen
  \bibfield  {author} {\bibinfo {author} {\bibfnamefont {Q.-X.}\ \bibnamefont
  {Lv}}, \bibinfo {author} {\bibfnamefont {Y.-X.}\ \bibnamefont {Du}}, \bibinfo
  {author} {\bibfnamefont {Z.-T.}\ \bibnamefont {Liang}}, \bibinfo {author}
  {\bibfnamefont {H.-Z.}\ \bibnamefont {Liu}}, \bibinfo {author} {\bibfnamefont
  {J.-H.}\ \bibnamefont {Liang}}, \bibinfo {author} {\bibfnamefont {L.-Q.}\
  \bibnamefont {Chen}}, \bibinfo {author} {\bibfnamefont {L.-M.}\ \bibnamefont
  {Zhou}}, \bibinfo {author} {\bibfnamefont {S.-C.}\ \bibnamefont {Zhang}},
  \bibinfo {author} {\bibfnamefont {D.-W.}\ \bibnamefont {Zhang}}, \bibinfo
  {author} {\bibfnamefont {B.-Q.}\ \bibnamefont {Ai}}, \bibinfo {author}
  {\bibfnamefont {H.}~\bibnamefont {Yan}},\ and\ \bibinfo {author}
  {\bibfnamefont {S.-L.}\ \bibnamefont {Zhu}},\ }\bibfield  {title} {\bibinfo
  {title} {Measurement of spin {C}hern numbers in quantum simulated topological
  insulators},\ }\href {https://doi.org/10.1103/PhysRevLett.127.136802}
  {\bibfield  {journal} {\bibinfo  {journal} {Phys. Rev. Lett.}\ }\textbf
  {\bibinfo {volume} {127}},\ \bibinfo {pages} {136802} (\bibinfo {year}
  {2021})}\BibitemShut {NoStop}%
\bibitem [{\citenamefont {Hao}\ and\ \citenamefont {Hu}(2014)}]{FeSe_PRX}%
  \BibitemOpen
  \bibfield  {author} {\bibinfo {author} {\bibfnamefont {N.}~\bibnamefont
  {Hao}}\ and\ \bibinfo {author} {\bibfnamefont {J.}~\bibnamefont {Hu}},\
  }\bibfield  {title} {\bibinfo {title} {Topological phases in the single-layer
  {FeSe}},\ }\href {https://doi.org/10.1103/PhysRevX.4.031053} {\bibfield
  {journal} {\bibinfo  {journal} {Phys. Rev. X}\ }\textbf {\bibinfo {volume}
  {4}},\ \bibinfo {pages} {031053} (\bibinfo {year} {2014})}\BibitemShut
  {NoStop}%
\bibitem [{\citenamefont {Luo}\ \emph {et~al.}(2022)\citenamefont {Luo},
  \citenamefont {Song},\ and\ \citenamefont {Xu}}]{FeSe_npj}%
  \BibitemOpen
  \bibfield  {author} {\bibinfo {author} {\bibfnamefont {A.}~\bibnamefont
  {Luo}}, \bibinfo {author} {\bibfnamefont {Z.}~\bibnamefont {Song}},\ and\
  \bibinfo {author} {\bibfnamefont {G.}~\bibnamefont {Xu}},\ }\bibfield
  {title} {\bibinfo {title} {Fragile topological band in the checkerboard
  antiferromagnetic monolayer {FeSe}},\ }\href
  {https://doi.org/10.1038/s41524-022-00707-9} {\bibfield  {journal} {\bibinfo
  {journal} {npj Comput. Mater.}\ }\textbf {\bibinfo {volume} {8}},\ \bibinfo
  {pages} {26} (\bibinfo {year} {2022})}\BibitemShut {NoStop}%
\end{thebibliography}%

\end{document}